\documentclass[10pt]{iopart}

\usepackage{iopams}
\usepackage[mathcal]{euscript}
\usepackage{amsgen,amsfonts,amssymb,amsbsy}

\usepackage{wasysym}
\usepackage{graphicx}% Include figure files
\usepackage{bm}% bold math
\usepackage[pdftex,bookmarks,colorlinks,citecolor=blue,linkcolor=blue]{hyperref}% add hypertext capabilities

\usepackage[left=1.0in,right=1.0in,top=1.0in,bottom=1.0in,includeheadfoot]{geometry}

\newenvironment{align}
 {\begin{eqnarray}}
 {\end{eqnarray}\ignorespacesafterend} 

\newcommand\eqref[1]{(\ref{#1})}

\newcommand\meantwo[1]{\left\langle #1 \right\rangle_2}

\newcommand{\D}{\mathrm{d}}
\newcommand\op[3]{\bm{#1}^{#2}_{#3}}

\newcommand\R[1]{\op{r}{}{#1}}

\newcommand{\rprim}{\bm{r}'}
\newcommand{\rdprim}{\bm{r}''}

\newcommand{\rs}{{r^*}}
\newcommand{\giso}{g_{\mathrm{iso}}}

\newcommand{\gd}{\dot{\gamma}}
\newcommand{\Uinf}{\bm{u}^{\infty}}
\newcommand{\UinfT}{\bm{u}^{\infty\,T}}
\newcommand{\DUinf}{\nabla\bm{u}^{\infty}}
\newcommand{\DUinfT}{\nabla\bm{u}^{\infty\,T}}
\newcommand{\Einf}{\bm{E}^{\infty}_{}}
\newcommand{\Ominf}{\bm{\Omega}^{\infty}_{}}

\newcommand{\gtwo}{g}
\newcommand{\gthree}{g_3}

\newcommand{\gSigma}{\bm{\Sigma}}
\newcommand{\gGamma}{\bm{\Gamma}}
\newcommand{\gXi}{\bm{\Xi}}
\newcommand{\gPi}{\bm{\Pi}}
\newcommand{\gUpsilon}{\bm{\Upsilon}}
\newcommand{\gTheta}{\bm{\Theta}}
\newcommand{\gLambda}{\bm{\Lambda}}
\newcommand{\gPhi}{\bm{\Phi}}
\newcommand{\gPsi}{\bm{\Psi}}

\newcommand{\gJ}{\bm{J}}

\newcommand{\gQ}{\bm{Q}}

\newcommand{\gF}{\bm{f}}

\newcommand{\gA}{\bm{A}}
\newcommand{\gB}{\bm{B}}

\newcommand{\ror}{\R{}\otimes\R{}}
\newcommand{\rof}{\R{}\otimes\gF(\R{})}
\newcommand\gdot{\bm{\cdot}}

\newcommand{\er}{\bm{e}_r}

\newcommand{\et}{\bm{e}_{\theta}}

\newcommand{\eone}{\bm{e}_1}
\newcommand{\etwo}{\bm{e}_2}
\newcommand{\eooeo}{\eone\otimes\eone}
\newcommand{\eooet}{\eone\otimes\etwo}
\newcommand{\etoeo}{\etwo\otimes\eone}
\newcommand{\etoet}{\etwo\otimes\etwo}
\newcommand{\eroer}{\er\otimes\er}

\newcommand\nablar{\nabla}

\def\identity{\;\mbox{l\hspace{-0.55em}1}}

\newcommand{\reveb}[1]{#1}

\newcommand{\comnc}[1]{\textcolor{black}{#1}}

\begin{document}

\title{Derivation of a constitutive model for the rheology of jammed soft suspensions from particle dynamics}

\author{Nicolas Cuny$^1$, Romain Mari$^1$, Eric Bertin$^1$}
\address{$1$ Univ. Grenoble Alpes, CNRS, LIPhy, 38000 Grenoble, France}

\date{\today}

\begin{abstract}
Considering the rheology of two-dimensional soft suspensions above the jamming density, we derive a tensorial constitutive model from the microscopic particle dynamics.
Starting from the equation governing the $N$-particle distribution, we derive an evolution equation for the stress tensor.
%using the Virial definition of stress. 
This evolution equation is not closed, as it involves the pair and three-particle correlation functions. 
To close this equation, we first employ the standard Kirkwood closure relation to express the three-particle correlation function in terms of the pair correlation function. Then we use a simple and physically motivated parametrization of the pair correlation function to obtain a closed evolution equation for the stress tensor.
The latter is naturally expressed as separate evolution equations for the pressure and for the deviatoric part of the stress tensor.
These evolution equations provide us with a non-linear tensorial constitutive model describing the rheological response of a jammed soft suspension to an arbitrary uniform deformation.
One of the advantages of this microscopically-rooted description is that the coefficients appearing in the constitutive model are known in terms of packing fraction and microscopic parameters.
\end{abstract}

\maketitle

%%%%%%%%%%%%%%%%%%%%%%%%%%%%%%%%%%%%%%%%%%%%%%%%%%%%%%%%%%%%%%%%%%%%%%%%%%%%%%%%%%%%%%%%%%%%%%%%%%%%%%%%%%%%%%%%%%%%%%%%%%%%%%%%%%%%%%%%%%%%%%%%%%%%%%%%%%%%%%%%%%%%%%%%%%%%%%%%%%%%%%%%%%%%%%%%%%%%%%%%%%%%%%%%%%

\section{\label{sec:intro}Introduction}

Microgel dispersions, emulsions, or suspensions of capsules are examples of suspensions of athermal soft particles within a fluid \cite{bonnecaze_micromechanics_2010}.
They have many current and potential applications in industry, 
% microgels
e.g. for drug delivery \cite{lopez_use_2005}, 
% microcapsules
thermal energy storage \cite{tyagi_development_2011}, 
% microcapsules
skin care products \cite{casanova_encapsulation_2016}, 
% microcapsules
food flavoring \cite{jun-xia_microencapsulation_2011}, 
% emulsions
or oil recovery \cite{peng_review_2017}.
% microcapsules in oil recovery abbaspourrad_polymer_2013
A large subset of these systems are essentially athermal, either because the microscopic constituents are too large, a micrometer and up, to be affected by Brownian motion, 
or because Brownian forces are dwarfed by other interactions at work, in particular in very concentrated systems for which elastic stresses from particle deformation 
are orders of magnitude larger than thermal stresses \cite{vlassopoulos_bridging_2012,vlassopoulos_tunable_2014}.
The latter case includes the common situation in applications for which the concentration (or volume fraction) of particles is large enough for the suspension to be jammed.
Above a jamming concentration $\phi_\mathrm{J}$, a yield stress $\sigma_\mathrm{y}$ proportional to the particles elastic modulus appears \cite{liu_jamming_1998,nordstrom_microfluidic_2010,liu_universality_2018}, leading to an elasto-plastic behavior as observed, e.g., in emulsions \cite{ma_rheological_1995} or gels \cite{roberts_new_2001}.
The material behaves essentially like an elastic solid under an applied stress $\sigma<\sigma_\mathrm{y}$, 
and flows for $\sigma>\sigma_\mathrm{y}$.

For such suspension, the steady state rheology 
\reveb{relating the shear stress $\sigma$ to the shear rate $\dot\gamma$}
under simple shear then follows a Herschel–Bulkley law $\sigma = \sigma_\mathrm{y} + k \dot\gamma^n$
with two material parameters, the consistency parameter $k$ and the flow parameter $n$.
In general, $n < 1$ is observed (the limiting case $n=1$ is called a Bingham fluid), and a typical value is $n\approx 0.5$ \cite{bonn_yield_2017,becu_yielding_2006,piau_carbopol_2007,seth_micromechanical_2011}.
This behavior is actually not restricted to the simple shear case, as it extends to general deformations, for which the rheology follows a tensorial HB law \cite{balmforth_yielding_2014}. 
The Herschel-Bulkley law is also observed in molecular dynamics simulations of particles interacting with soft potentials \cite{otsuki_universal_2009,otsuki_critical_2011,olsson_herschel-bulkley_2012,kawasaki_diverging_2015}, 
as well as in system-specific simulations using Discrete Element Method for microgel suspensions \cite{seth_micromechanical_2011}, 
or Immersed Boundary Method for suspensions of capsules \cite{gross_rheology_2014}.
The transient rheology of soft particles suspensions also displays non-trivial features, like an overshoot when a sample initially at rest is brought to yield \cite{coussot_macroscopic_2009,divoux_stress_2011,dinkgreve_different_2016}, or a relaxation to a smaller final stress after a faster preshear \cite{mohan_microscopic_2013,mohan_build-up_2014}.

On the theory side, there is currently no system-specific approaches to the rheology of jammed soft athermal suspensions.
There are however many theoretical approaches to the rheology of yield stress fluids (YSF) in general, which consists of a much wider class of systems 
including soft jammed suspensions as well as colloidal gels or glasses \cite{bonn_yield_2017}.
By design, these approaches are agnostic regarding the microscopic interactions giving rise to the stress, 
and thereby to the yield stress.
As a consequence, the yield stress is often (and even always when it comes to describing athermal materials) 
an input of these approaches, not an outcome of some more fundamental description.

At the macroscopic level, constitutive models postulate stress evolution equations built around the existence of a yield stress.
The domain of acceptable constitutive models is mostly bounded by symmetries of the underlying problem and thermodynamic laws. 
%For YSF, we can distinguish two main classes. 
%Viscoplastic models ignore the elasticity of the suspension before yield \cite{papanastasiou_flows_1987}: the suspension is then either a rigid solid for $\sigma<\sigma_\mathrm{y}$ or a visco-plastic material for $\sigma>\sigma_\mathrm{y}$. 
%On the contrary, elasto-viscoplastic models consider that the material has an elastic response before yielding \cite{yoshimura_response_1987,doraiswamy_coxmerz_1991,saramito_new_2007,saramito_new_2009,benito_elasto-visco-plastic_2008,park_oscillatory_2010,belblidia_computations_2011,dimitriou_describing_2012,dimitriou_canonical_2019}.
These models can be designed to exhibit a steady-state rheology following either Bingham law \cite{saramito_new_2007,benito_elasto-visco-plastic_2008} 
or the more general Herschel-Bulkley law \cite{saramito_new_2009,belblidia_computations_2011,dimitriou_describing_2012,kamaniUnificationRheologicalPhysics2021}.
Additionally, they may include spatial cooperativity via non-local terms \cite{goyon_spatial_2008}.
These models can achieve quantitative agreement with experiments in steady flow in nontrivial geometries \cite{goyon_spatial_2008,cheddadi_understanding_2011} or in time-dependent flows \cite{dimitriou_describing_2012,dimitriou_canonical_2019}.
As phenomenological models, however, the choice of their structure is driven by the behavior they are intended to describe, 
and their parameters are usually free parameters, which values are fixed by fits to reference properties of the material 
to be modelled.
The connection to microscopic properties of particles, like their stiffness or interaction laws, 
is only qualitative: for instance, other things held constant, 
suspensions of stiffer particles should be described by models predicting a steeper elastic branch in the load curve.

A partial attempt to connect macroscopic properties to the underlying physics is given by mesoscopic models such as the Soft Glassy Rheology (SGR) model \cite{sollich_rheology_1997,sollich_rheological_1998}, elasto-plastic models \cite{hebraud_mode-coupling_1998,nicolas_rheology_2014,lin_scaling_2014,nicolas_deformation_2018}, or Shear Transformation Zone (STZ) theory \cite{falk_dynamics_1998,bouchbinder_athermal_2007}.
These models have in common to consider the emergence of nontrivial macroscopic phenomena (e.g., a Herschel-Bulkley rheology at low shear rate) from the statistics of a large assembly 
of mesoscopic building blocks (typically mesoscopic plastic or elasto-plastic elements) which do not individually possess these nontrivial features. 
In spite of their success in predicting flow curves \cite{nicolas_rheology_2014,lin_scaling_2014} or transient rheology  \cite{bouchbinder_athermal_2007,picard_slow_2005,lin_criticality_2015}, such mesoscopic models do not establish a connection between  microscopic (particle-level) properties and the mechanical behavior of the mesoscopic elements, 
which are given some simple mechanical properties relating inherently macroscopic quantities, stress and deformation.

So far, microscopic approaches to the rheology of YSF have been restricted to colloidal glasses.
The main microscopic approach for these systems is Mode-Coupling Theory (MCT), in particular through its Integration Through Transients (ITT) variant \cite{fuchs_theory_2002,fuchs_schematic_2003,fuchs_integration_2005,fuchs_mode_2009,brader_glass_2009}, which was used to address steady-state features as well as transient 
ones like stress overshoots and stress relaxation \cite{amann_overshoots_2012,ballauff_residual_2013}.
MCT is suited to describe systems close to thermodynamic equilibrium for which the yield stress stems from the glass transition, not the jamming transition, 
and the rheology is a competition between shear forces creating anisotropy 
and Brownian forces restoring stationary isotropy. 
Recent extensions have however been introduced to address granular material rheology \cite{coquand_integration_2020}.

In this article, we detail a novel approach to the modeling of the rheology of jammed athermal suspensions of soft particles. 
In this approach, we derive a constitutive model directly from the microscopic dynamics of a model bidimensional suspension.
Starting from a time evolution equation on the pair correlation function, we obtain an evolution equation on the stress tensor. However, this equation is not closed both because it involves the three-body correlation function, and because even terms involving only the pair correlation function are not directly expressed in terms of the stress tensor. To close the evolution equation of the stress tensor, we first use the Kirkwood superposition approximation to express the three-body correlation function as a product of pair correlation functions. Second, we perform a weakly anisotropic expansion of the pair correlation function to express it in terms of isotropic pair correlation function and of a structure tensor that encodes the anistropy. Finally, a simple and physically motivated parameterization of the isotropic pair correlation function allows us to get a closed form for the evolution equation of the stress tensor.
The resulting constitutive model takes the form of a pair of coupled ODEs
for the time evolution of respectively the deviatoric part of the particle stress tensor and its trace (the particle pressure)
as a function of the applied deformation rate tensor.
The parameters of this model can be directly related to particle properties.
The methodology presented here can be systematically generalized to addressed other geometries (e.g., three-dimensional systems) or physical regimes like dense suspensions just below the jamming transition, when hydrodynamic interactions remain screened and elastic interactions still play an important role. In addition, the present methodology may also be extended to other types of athermal systems like dense soft active systems (e.g., epithelial tissues \cite{RanftFluidization2010,matozNonlinear2017}), in order to obtain constitutive relations for their rheology.

In its principle, this approach is closely related to approaches for deriving constitutive relations for polymeric systems, 
starting from the Smoluchowski equation \cite{doi_theory_1988}, with however key differences.
The jammed suspensions we consider are athermal, unaffected by diffusion and thus arbitrarily far from equilibrium.
Moreover, contrary to polymers in concentrated solutions or in melts, suspended particles interact with a small number of their neighbors (typically $\apprle 10$), 
so that mean-field approximations to the dynamics are not expected to be accurate.
Previous works have followed the Smoluchowski equation route in the field of suspensions \cite{lionberger_smoluchowski_1997,nazockdast_microstructural_2012,banetta_pair_2020}. 
These works focused on colloidal suspensions below jamming, where Brownian motion and hydrodynamic interactions are significant contributors to the stress response.
In contrast, the present work focuses on jammed suspensions, for which the elastic deformation of particles is dominant \cite{liu_universality_2018}.

The article is organized as follows. In Sec.~\ref{sec:model}, we introduce our microscopic model of a suspension of soft particles, 
for which we will derive a constitutive model.
Sections~\ref{sec:level1}--\ref{sec:calculation_coefficient} are dedicated to the progressive derivation of the constitutive model. 
Sec.~\ref{sec:level1} derives the exact (but not closed) dynamics of the pair correlation function. 
From this result, we derive an equally exact (and equally not closed) dynamics for the stress tensor in Sec~\ref{sec:eq_sigma}.
Then, in Sec.~\ref{sec:closure} and~\ref{sec:calculation_coefficient}, we perform a closure of the stress dynamics, based on physically motivated approximations 
for the pair correlation function. The final constitutive model, given in Eqs.~(\ref{eq_sigmaprim_closed_p}) and (\ref{eq_p_closed_p}), takes the form of non-linear, coupled evolution equations for the pressure and the deviatoric part of the stress tensor.
Finally we discuss our approach, its limitations and possible future developments in Sec.~\ref{sec:discussion}.

%\input{model}

%!TEX root = article.tex

\section{\label{sec:level1}Model and microstructure dynamics}

%!TEX root = article.tex

\subsection{\label{sec:model}Soft suspension model}

We consider a two-dimensional system of $N$ athermal, overdamped, circular particles with the same radius $a$. Particles are not subjected to gravity and interact one with another 
by radial contact repulsion forces only. 
They are immersed in a Newtonian fluid generating a viscous drag on them;
however particles are assumed not to act on the fluid.
Consequently, the fluid is described by an affine velocity field $\bm{u}^{\infty}(\bm{r})$ unaffected by the particle dynamics. 
The gradient of this velocity field is assumed uniform, that is $\bm{u}^{\infty}(\bm{r}) = \nabla \bm{u}^{\infty} \gdot \bm{r}$
(we define the velocity gradient as $(\nabla\mathbf{u}^{\infty})_{ij}=\partial \mathbf{u}^{\infty}_i/\partial r_j$).
The system has a volume $V$, and a particle density $\rho=N/V$, which we also assume uniform. 

The position of particle $\mu$ is $\op{r}{}{\mu}$, and its velocity $\dot{\bm{r}}_\mu$. 
We call $\op{u}{\infty}{\mu}=\bm{u}^{\infty}(\bm{r}_{\mu})$ the fluid velocity field at the position of particle $\mu$. The viscous drag acting on particle $\mu$ is then $-\lambda_{\rm f} (\dot{\bm{r}}_\mu-\op{u}{\infty}{\mu})$.
Particles interact through pairwise repulsive contact forces. 
The repulsive force applied by particle $\nu$ on particle $\mu$ is 
$\bm{f}(\op{r}{}{\mu\nu}) = f(r_{\mu\nu})\op{r}{}{\mu\nu}/r_{\mu\nu}$, with $\op{r}{}{\mu\nu}=\op{r}{}{\nu}-\op{r}{}{\mu}$, $r_{\mu\nu} = |\op{r}{}{\mu\nu}|$ (note that $f(r) \le 0$ for the repulsive force considered here).
We keep the contact model generic for the moment, with the only requirement that $f(r_{\mu\nu}) = 0$ for $r_{\mu\nu} > 2a$.
Calling $f_0$ a typical contact force, we define a unit system with $f_0$ the unit force, $\tau_0=\lambda_{\rm f} a/(2f_0)$ the unit time, and $a$ the unit length. 
In most cases of interest, $f_0$ can be defined from the expansion $f(r) \sim -f_0 [(2a-r)/a]^{\alpha_{\rm f}}$ at contact,
$r \to 2a^{-}$ (with $\alpha_{\rm f}>0$ a model-dependent exponent).
Briefly introducing the notation $\hat{X}$ for the dimensionless $X$ ($X$ being any physical variable), the dimensionless equation of motion for particle $\mu$ then reads
\begin{equation}
	\label{dynamics}
	-2 (\hat{\dot{\bm{r}}}_\mu - \hat{\bm{u}}^{\infty}_{\mu})+\hat{\bm{f}}_{\mu} = \bm{0},
\end{equation}
with $\hat{\bm{f}}_{\mu} = \sum_{\nu\neq \mu} \hat{\bm{f}}(\hat{\bm{r}}_{\mu\nu})$ the resulting contact force on particle $\mu$.
In the following, we will work with dimensionless variables, 
and for the sake of readability drop the $\hat{X}$ notation.

In this section, we derive an exact equation on the pair correlation function of the suspension. 
This part of the derivation parallels other works in the literature (for instance~\cite{nazockdast_microstructural_2012}), but we reproduce it here for the present article to be self-contained.

\subsection{\label{subsec:probability_conservation}Conservation of probability}

A $N$-body probability function $P_N$ is associated with the system. $P_N$ is a function of $N$ position vectors such that $P_N(\R{1},\ldots,\R{N})$ 
is the probability to find the $N$ particles at the respective positions $\R{1},\ldots,\R{N}$. The conservation equation characterizing the evolution of this function is
\begin{equation}
	\label{probability_conservation}
	\partial_t P_N + \sum\limits_{\mu=1}^{N}\nabla_{\mu} \gdot \op{j}{}{\mu}=0,
\end{equation}
with $\op{j}{}{\mu}$ the probability current vector associated with particle $\mu$, defined as
\begin{equation}
	\label{j_alpha}
	\op{j}{}{\mu}=\left(\op{u}{\infty}{\mu} + \frac{1}{2}\op{f}{}{\mu}\right)P_N.
\end{equation}
Note that we use here a probabilistic description although the dynamics is purely deterministic due to the absence of thermal noise. Hence here the probabilistic description does not account for stochasticity, but for the deterministic (dissipative) evolution of a set of initial condition, in the same spirit as, e.g., the Liouville equation for Hamiltonian systems.

Integrating $P_N$ over $N-k$ position vectors, we obtain the $k$-body reduced probability distribution $\mathcal{P}_k$,
\begin{equation}
	\label{def_Pk}
	\mathcal{P}_k(\R{1},\ldots,\R{k})=\frac{N!}{(N-k)!}\int 
%	P_N(\R{1},\ldots,\R{N})
	P_N(\{\R{j}\}) \D\R{k+1}\ldots\D\R{N},
\end{equation}
where \reveb{the notation $\D\R{i}$ stands for the two-dimensional integration element over the variable $\R{i}$.}
The combinatorial factor $N!/(N-k)!$ takes into account the fact that all particles play an equivalent role
(i.e., the chosen particles $\{1,\dots,k\}$ could be any subset of $k$ particles in the system). 

In particular, we reduce Eq.~(\ref{probability_conservation}), which is a $2N$-variable equation,
to a 4-variable equation by integrating it over $N-2$ position vectors to get 
\begin{equation}
	\label{eq1}
	\frac{\partial_t \mathcal{P}_2(\R{1},\R{2})}{N(N-1)} + \int \nabla_{1} \!\gdot\! \op{j}{}{1}\D \op{r}{}{3}...\D \op{r}{}{N}
	+ \int \nabla_{2} \!\gdot\! \op{j}{}{2} \D \op{r}{}{3}...\D \op{r}{}{N} %\nonumber\\ 
	+ \sum\limits_{\mu=3}^{N} \int \nabla_{\mu} \!\gdot\! \op{j}{}{\mu} \D \op{r}{}{3}...\D \op{r}{}{N} = 0.
\end{equation}
Integrals in the sum can be calculated applying the Green-Ostrogradski theorem, which gives
\begin{equation}
\label{green-ostrogradski}
	\int \! \nabla_{\mu} \! \gdot \op{j}{}{\mu} \D \op{r}{}{3}...\D \op{r}{}{N} \!=\! \oint\limits_{\partial V} \op{j}{}{\mu} \! \gdot \D \mathbf{S}_\mu \D \op{r}{}{3}...\D \op{r}{}{\mu-1} \D \op{r}{}{\mu+1}...\D \op{r}{}{N} = 0,
\end{equation}
thanks to the no-flux boundary conditions.

We also introduce $\op{J}{}{\mu}(\R{1}{},\R{2}{})=\int \op{j}{}{\mu}\D \op{r}{}{3}...\D \op{r}{}{N}$. Assuming that the material is homogeneous such that $P_2$, $\op{J}{}{1}$ and $\op{J}{}{2}$ are only functions of $\R{}=\R{2}-\R{1}$, we reduce to two the number of (real) variables. We also notice that $\nabla_{2}=-\nabla_{1} \equiv \nablar$ (where $\nablar$ stands for the gradient with respect to the relative position $\R{}$), leading to

\begin{equation}
	\label{eq_P2}
	\partial_t P_2(\R{}) + N(N-1) \nablar \gdot \op{J}{}{12} =0,
\end{equation}
with vector $\op{J}{}{12}$ defined as
\begin{equation}
	\label{def_J12}
	\op{J}{}{12} =\op{J}{}{2}-\op{J}{}{1} =\int\left( \op{u}{\infty}{12} + \op{f}{(12)}{} \right) P_N(\R{1}{},...,\R{N}{}) \D \op{r}{}{3}...\D \op{r}{}{N},
\end{equation}
where $\op{u}{\infty}{12}=\op{u}{\infty}{2}-\op{u}{\infty}{1}$ and $\op{f}{(12)}{}=\frac{1}{2}\left(\op{f}{}{2}-\op{f}{}{1}\right)$. 

\subsection{\label{subsec:eq_g}Dynamics of the pair correlation function}

We then introduce the notation $\meantwo{.}$ representing the mean value of an observable over the
$(N-2)$-particle configuration space,
\begin{align}
	\label{mean2}
	\meantwo{A}\!(\R{}) & = \frac{\int \! A(\R{},\R{3},\ldots,\R{N}) P_N(\R{},\R{3},\ldots,\R{N}) \D \op{r}{}{3}...\D \op{r}{}{N}}{\int P_N(\R{},\R{3},\ldots,\R{N}) \D \op{r}{}{3}...\D \op{r}{}{N}} \nonumber\\
					  & = \frac{N(N-1)}{P_2(\R{})} \int A(\R{},\R{3},\ldots,\R{N}) P_N(\R{},\R{3},\ldots,\R{N}) \D \op{r}{}{3}...\D \op{r}{}{N}. \nonumber\\
\end{align}
Introducing also the pair correlation function $\gtwo(\R{})=P_2(\R{})/\rho^2$, we can rewrite Eq.~\eqref{eq_P2} as

\begin{equation}
\label{eq_mean2}
\partial_t \gtwo(\R{}) = - \nablar \gdot \left( \op{u}{\infty}{12}(\R{}) \gtwo(\R{}) + \meantwo{\op{f}{(12)}{}}\!(\R{})\, \gtwo(\R{})\right),
\end{equation}
where we used the fact that $\meantwo{\op{u}{\infty}{12}}=\op{u}{\infty}{12}$.

% \subsubsection{\label{subsubsec:pairwise_additivity}Pairwise additivity of the contact force}

Using the pairwise additivity of contact forces we can write
\begin{equation}
	\label{pairwise_additivity1}
	\meantwo{\op{f}{}{1}}\gtwo(\R{})  = 
						 \gF(\R{})\gtwo(\R{})+\rho \int \gF(\rprim)\gthree(\R{},\rprim) \D \rprim,
\end{equation}
where $\rprim$ is the center-to-center vector between particles 1 and 3 and $\gthree$ the three-body correlation function reduced to four variables using translational invariance,
\begin{equation}
\gthree(\R{},\rprim) = \frac{1}{\rho^3} \, \mathcal{P}_3(\mathbf{0},\R{},\rprim)\,.
\end{equation}
Similarly, for particle 2
\begin{align}
	\label{pairwise_additivity2}
	\meantwo{\gF_2}\gtwo(\R{}) & = \gF(-\R{})\gtwo(\R{})+\rho \int \gF(\rprim-\R{}) \gthree(\R{},\rprim) \D \rprim \nonumber\\
						   & = -\gF(\R{})\gtwo(\R{})+\rho \int \gF(\rprim-\R{}) \gthree(\R{},\rprim) \D \rprim.
\end{align}
All particles playing an equivalent role, $\gthree$ satisfies some symmetries resulting from particle permutations. In particular, the permutation of particle 1 with particle 2 implies that $\gthree(\R{},\rprim)=\gthree(-\R{},\rprim-\R{})$. The fact that $\Uinf$ and $\gF$ are odd under the change $\R{} \to -\R{}$ also implies $\gthree(\R{},\rprim)=\gthree(-\R{},-\rprim)$. We can thus show that
\begin{align}
	\label{pairwise2_int_step}
	\int \gF(\rprim-\R{}) \gthree(\R{},\rprim) \D \rprim & = \int \gF(\rprim-\R{}) \gthree(-\R{},\rprim-\R{}) \D \rprim \nonumber\\
													 & = \int \gF(\rdprim) \gthree(-\R{},\rdprim) \D \rdprim \nonumber\\
													 & = \int -\gF(-\rdprim) \gthree(\R{},-\rdprim) \D \rdprim \nonumber\\
													 & = -\int \gF(\rprim) \gthree(\R{},\rprim) \D \rprim .
\end{align}
Using this result in Eq.~(\ref{pairwise_additivity2}), we obtain that $\meantwo{\gF_2}=-\meantwo{\gF_1}$, and thus $\meantwo{\gF^{(12)}}=\meantwo{\gF_2}$. Replacing $\Uinf_{12}$ and $\meantwo{\gF^{(12)}}$ by their respective expressions in Eq.~(\ref{eq_mean2}), we get
\begin{equation}
	\label{eq_g2}
	\partial_t \gtwo(\R{}) + \mathbf{\nabla} \gdot \,\gJ(\R{}) = 0,
\end{equation}
with
\begin{equation}
	\label{def_J}
	\gJ(\R{})= \DUinf \gdot \R{}\,\gtwo(\R{}) -\gF(\R{})\gtwo(\R{})-\rho \int \gF(\rprim) \gthree(\R{},\rprim) \D \rprim.
\end{equation}
We thus get an exact dynamical equation on $g$, however not closed because of the presence of $\gthree$.

%%%%%%%%%%%%%%%%%%%%%%%%%%%%%%%%%%%%%%%%%%%%%%%%%%%%%%%%%%%%%%%%%%%%%%%%%%%%%%%%%%%%%%%%%%%%%%%%%%%%%%%%%%%%%%%%%%%%%%%%%%%%%%%%%%%%%%%%%%%%%%%%%%%%%%%%%%%%%%%%%%%%%%%%%%%%%%%%%%%%%%%%%%%%%%%%%%%%%%%%%%%%%%%%%%

%!TEX root = article.tex

\section{\label{sec:eq_sigma}Exact stress tensor dynamics}

Our aim is to obtain an evolution equation for $\gSigma$, the elastic part of the stress tensor of the material. This macroscopic quantity is intimately linked to the micro-structure of the material through the Virial formula \cite{nicot_definition_2013}
\begin{equation}
	\label{virial}
	\gSigma=\frac{\rho^2}{2}\int\left(\rof\right) g(\R{})\D\R{}.
\end{equation}
Hence, multipliying Eq.~(\ref{eq_g2}) by $\frac{1}{2}\rho^2 \left(\rof\right)$ and integrating on $\R{}$, we get the time evolution equation for the stress

\begin{equation}
	\label{eq_sigma_J}
	\dot{\gSigma} + \frac{\rho^2}{2}\int \left(\rof\right) \nabla \gdot \,\gJ(\R{}) \D\R{}=\mathbf{0}.
\end{equation}
We then make use of the Green-Ostrogradski relation to reformulate the integral term as
\begin{equation}
	\label{dvpt_intdjsigma_tenseur}
	\int (\rof) \nabla \gdot \gJ(\R{}) \D\R{} = -\int \gJ\otimes\gF\D\R{} - \int(\R{}\otimes\gJ)\,\gdot\,\nabla\gF^T \D\R{},
\end{equation}
where we also used the fact that $\gF$ vanishes for non-contacting particles, which leads to a vanishing boundary term.

Replacing $\gJ$ by its expression Eq.~(\ref{def_J}), we thus get:

\begin{equation}
	\label{eq_sigma}
	\dot{\gSigma} =\DUinf \gdot\, \gSigma + \mathbf{\Lambda} - \mathbf{\Xi} - \mathbf{\Pi} - \mathbf{\Gamma} - \mathbf{\Upsilon},
\end{equation}
with
\begin{align}
	\label{def_Lambda}
	&\mathbf{\Lambda}=\frac{\rho^2}{2} \int (\ror)\left(\nabla\gF\,\gdot\,\nabla\Uinf\right)^T g(\R{})\D\R{}, \\
	\label{def_Xi}
	&\mathbf{\Xi}= \frac{\rho^2}{2} \int \left(\gF(\R{})\otimes\gF(\R{})\right)g(\R{})\D\R{}, \\
	\label{def_Pi}
	&\mathbf{\Pi}=\frac{\rho^2}{2} \int\left(\R{}\otimes\gF(\R{})\right)\gdot\,\nabla\gF\, g(\R{})\D\R{},\\
	\label{def_Gamma}
	&\mathbf{\Gamma}=\frac{\rho^3}{2}\iint \left(\gF(\rprim)\otimes \gF(\R{})\right) \gthree(\R{},\rprim)\D\R{}\D\rprim, \\
	\label{def_Upsilon}
	&\mathbf{\Upsilon}=\frac{\rho^3}{2} \iint \left(\R{} \otimes \gF(\rprim)\right)\gdot \nabla\gF(\R{})\,\gthree(\R{},\rprim)\D\R{}\D\rprim,
\end{align}
where we have used the fact that $\nabla\gF$ is a symmetric tensor for a radial force $\gF$. Using again this last assumption, we can moreover show that (details of the calculation can be found in \ref{app:calc_lambda})
\begin{align}
	\label{dvpt_lambda_tens}
	\mathbf{\Lambda}  = \gSigma \gdot \nabla\UinfT & + \frac{\rho^2}{2} \int \left(\Einf:\eroer\right) (\ror)\gdot\nabla\gF\, g(\R{})\D\R{}  \nonumber\\
					 & - \frac{\rho^2}{2} \int \left(\Einf:\eroer\right) (\rof) g(\R{})\D\R{},
\end{align}
where we introduced $\er=\R{}/|\R{}|$ as well as the strain-rate tensor $\Einf$, defined as the symmetric part of $\DUinf$,
\begin{equation}
\Einf = \frac{1}{2} (\DUinf + \DUinfT)\,.
\end{equation}
\reveb{The double dot product of two tensors $A$ and $B$ is defined as the scalar $A\!:\!B=\sum_{i,j} A_{ij}B_{ij}$.}
Inserting Eq.~(\ref{dvpt_lambda_tens}) in Eq.~(\ref{eq_sigma}), we get
\begin{equation}
	\label{eq_sigma_with_Theta}
	\dot{\gSigma} = \DUinf \gdot\, \gSigma + \gSigma \gdot\, \DUinfT + \mathbf{\Theta} - \mathbf{\Phi} - \mathbf{\Xi} - \mathbf{\Pi} - \mathbf{\Gamma} - \mathbf{\Upsilon},
\end{equation}
with
\begin{align}
	\label{def_Theta}
	&\mathbf{\Theta} = \frac{\rho^2}{2} \int \left(\Einf:\eroer\right) (\ror)\gdot\nabla\gF\, g(\R{})\D\R{}, \\
	\label{def_Phi}
	&\mathbf{\Phi} = \frac{\rho^2}{2} \int \left(\Einf:\eroer\right) (\rof)\, g(\R{})\D\R{}.
\end{align}

We can decompose this tensorial equation in a traceless equation coupled with a scalar equation on the trace. 
From now on, we denote as $\gA'$ the deviatoric part of any symmetric tensor $\gA$, that is
 $\gA'=\gA - \frac{1}{2} (\tr\gA)\identity$ (in two dimensions). In particular, for $\DUinf \gdot\, \gSigma + \gSigma \gdot\, {\DUinf}^T$ we get
\begin{align}
	\label{tr_duinfsigma}
	&\tr\left(\DUinf \gdot\, \gSigma + \gSigma \gdot\, {\DUinf}^T\right) = 2\left(\Einf:\gSigma'\right), \\
	\label{deviatoric_duinfsigma}
	&\left(\DUinf \gdot\, \gSigma + \gSigma \gdot\, {\DUinf}^T\right)' = 
	 \Ominf\gdot\,\gSigma'-\gSigma'\gdot\,\Ominf + \left(\tr\gSigma\right)\Einf,
\end{align}
with $\Ominf$ the antisymmetric part of $\DUinf$, 
\begin{equation}
\Ominf = \frac{1}{2} (\DUinf - \DUinfT)\,.
\end{equation}
also called the vorticity tensor.
Details of the derivation can be found in \ref{app:calc_lambda}. 
An important aspect is that the result (\ref{deviatoric_duinfsigma}) is specific to bidimensional systems because $\Einf\gdot\gSigma'+\gSigma'\gdot\Einf=(\Einf:\gSigma')\identity$ only in two dimensions.

Introducing the pressure $p=-\frac{1}{2}\tr\gSigma$, we get the following decomposition for 
the dynamics of the particle stress tensor, with a time evolution of the deviatoric part of the stress
\begin{equation}
	\label{eq_sigmaprim}
	\dot{\gSigma}'  = \Ominf \gdot\, \gSigma' - \gSigma' \gdot\, \Ominf -2p\:\Einf  + \mathbf{\Theta}' - \mathbf{\Phi}' - \mathbf{\Xi}' - \mathbf{\Pi}' - \mathbf{\Gamma}' - \mathbf{\Upsilon}',
\end{equation}
and another evolution equation for the particle pressure
\begin{equation}
	\label{eq_p}
	\dot{p} = -\left(\gSigma':\Einf\right) - \frac{\tr\mathbf{\Theta}}{2} + \frac{\tr\mathbf{\Phi}}{2} + \frac{\tr\mathbf{\Xi}}{2} + \frac{\tr\mathbf{\Pi}}{2} + \frac{\tr\mathbf{\Gamma}}{2} + \frac{\tr\mathbf{\Upsilon}}{2}.
\end{equation}
We thus get exact macroscopic evolution equations for the deviatoric particle stress and the particle pressure. However, these are not yet closed, as the tensors appearing in these equations are not directly expressed in term of $\gSigma'$ and $p$.
The closure of these evolution equations is the purpose of the next sections.
%%%%%%%%%%%%%%%%%%%%%%%%%%%%%%%%%%%%%%%%%%%%%%%%%%%%%%%%%%%%%%%%%%%%%%%%%%%%%%%%%%%%%%%%%%%%%%%%%%%%%%%%%%%%%%%%%%%%%%%%%%%%%%%%%%%%%%%%%%%%%%%%%%%%%%%%%%%%%%%%%%%%%%%%%%%%%%%%%%%%%%%%%%%%%%%%%%%%%%%%%%%%%%%%%%

%!TEX root = article.tex
\section{\label{sec:closure}Closure of the stress dynamics}

Up to this point all our calculations are exact with respect to the microscopic model we chose. 
This led us to a pair of unclosed equations on the deviatoric particle stress tensor, Eq.~(\ref{eq_sigmaprim}), 
and on the particle pressure, Eq.~(\ref{eq_p}). 
To close these equations, we have to introduce some approximations in the description of the microstructure.
In order to test some of these approximations, we compare them with the results of \reveb{numerical simulations that we briefly describe below.}

\subsection{\comnc{Molecular Dynamics simulations}}

\reveb{Numerical simulations have been performed using}
LAMMPS~\cite{plimptonFastParallelAlgorithms1995}, on a 2D system 
of $1000$ disks interacting through a harmonic \reveb{repulsive} potential~\cite{durian_foam_1995}, under simple shear $(\nabla \bm{u}^\infty)_{ij} = \dot\gamma \delta_{ix} \delta_{jy}$.
\comnc{The disks follow the dynamics of our model, defined in Eq.~\eqref{dynamics}, with $f(r)=r-2$ if $r<2$ and $f(r)=0$ \reveb{otherwise}. Boundary conditions are periodic, using Lees-Edwards conditions along the gradient direction~\cite{leesComputerStudyTransport1972}}.
To avoid strong \reveb{crystal} ordering, we use a $50:50$ bidisperse mixture 
with size ratio $1:1.4$~\cite{speedyGlassTransitionHard1999}.
\comnc{We determine numerically the stationary state of our system under simple shear, starting from different initial configurations so as to obtain 300 independent realisations of the system in stationary state for each value of density and shear rate.}
To allow for comparison with our monodisperse theory, we define the numerical correlation functions as 
\begin{align}
    &g(\bm{r}) =\frac{1}{\rho N}\sum_{i\neq j} \delta(\bm{r} - \tilde{\bm{r}}_{ij}) \\
    &\gthree(\bm{r}, \bm{r}')  = \frac{1}{\rho^2 N}\sum_{i\neq j \neq k} \delta(\bm{r}' - \tilde{\bm{r}}_{ij}) \delta(\bm{r} - \tilde{\bm{r}}_{ik}),
\end{align}
with \reveb{appropriately rescaled interparticle distances} $\tilde{\bm{r}}_{ij}=2(\bm{r}_j -\bm{r}_i)/(a_i+a_j)$ and $a_i$ the radius of particle $i$. \comnc{Those quantities are obtained by averaging over the 300 configurations.}

\subsection{\label{subsec:affine_approx_g}Weak anisotropy}

\begin{figure}
	\begin{center}
		\includegraphics[width=0.5\linewidth]{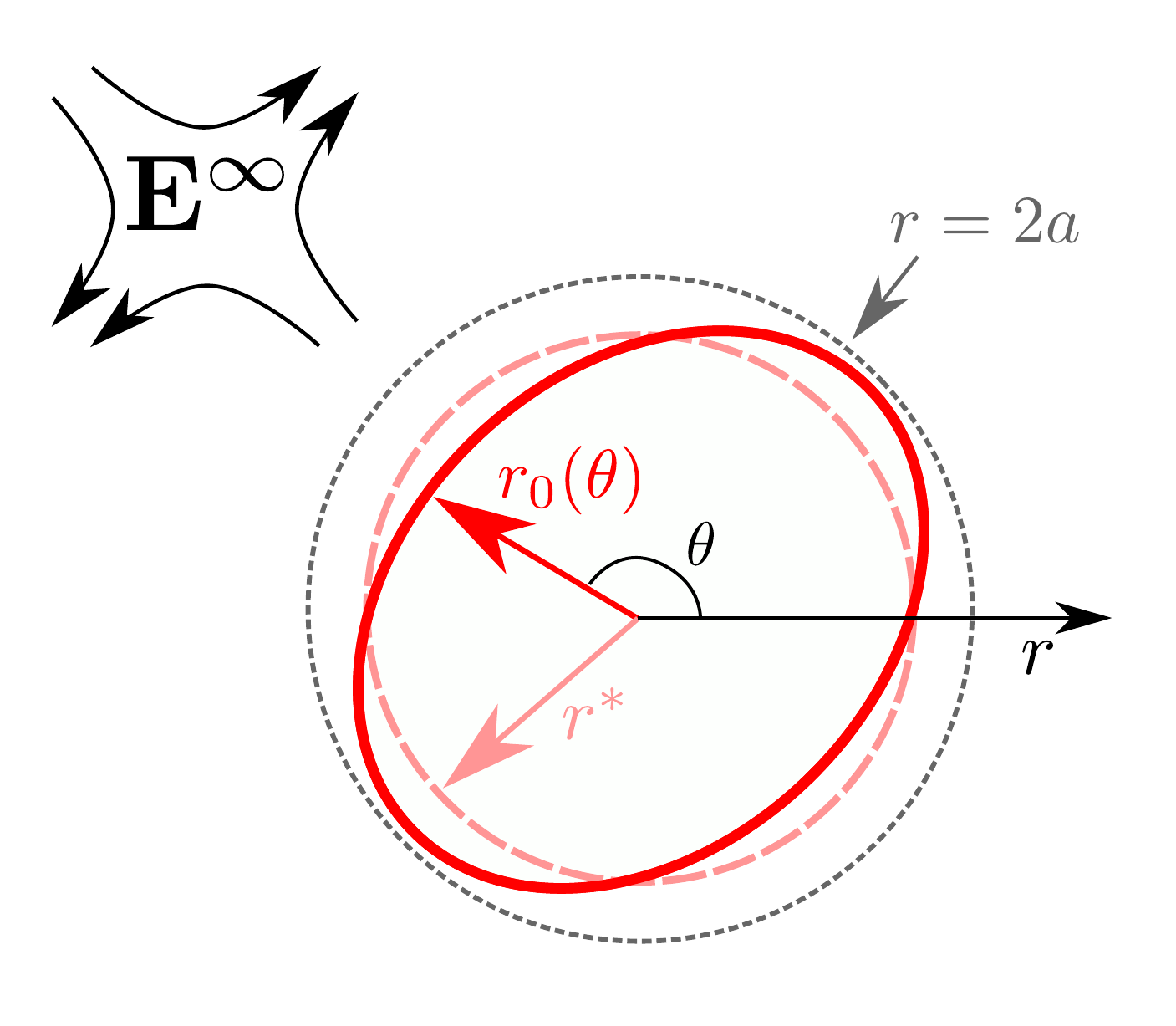}
		\caption{Illustration of the deformed $\giso(r)$ hypothesis. We assume that the pair correlation function $g(\bm{r})$, here only represented by a sketch of its first-neighbor peak in polar coordinates $r,\theta$ (solid red line), is a deformation of the isotropic $\giso(r)$ (which first-neighbor peak is the pink dashed line) under the action of the imposed deformation rate tensor $\Einf$.}  
		\label{fig:ellipse}
	\end{center}
\end{figure}

From the definition of the particle stress in Eq.~(\ref{virial}), it is clear that 
the deviatoric part of the stress is borne from the anisotropy of the microstructure.
The simplest measure of this anisotropy is the so-called structure tensor
\begin{equation}
    \label{eq:def:Q}
    \gQ=\frac{\rho^2}{2}\int_{|\R{}|\le2}\left[\ror - \frac{|\R{}|^2}{2} \identity\right]g(\R{})\D\R{}\,,
\end{equation}
which is by definition a traceless tensor.
\begin{figure}
	\begin{center}
		\includegraphics[height=5.6cm]{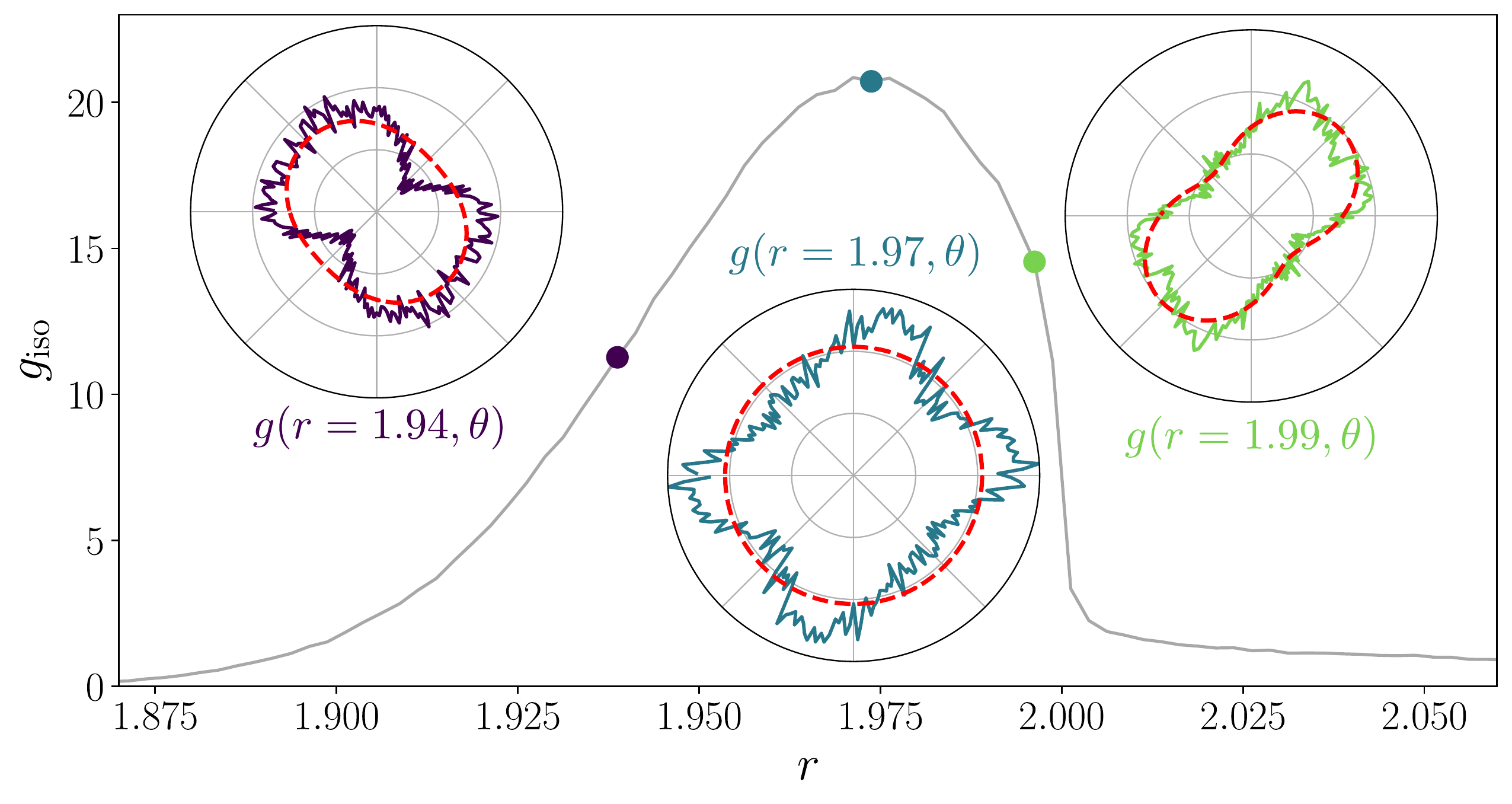}
		\includegraphics[height=5.6cm]{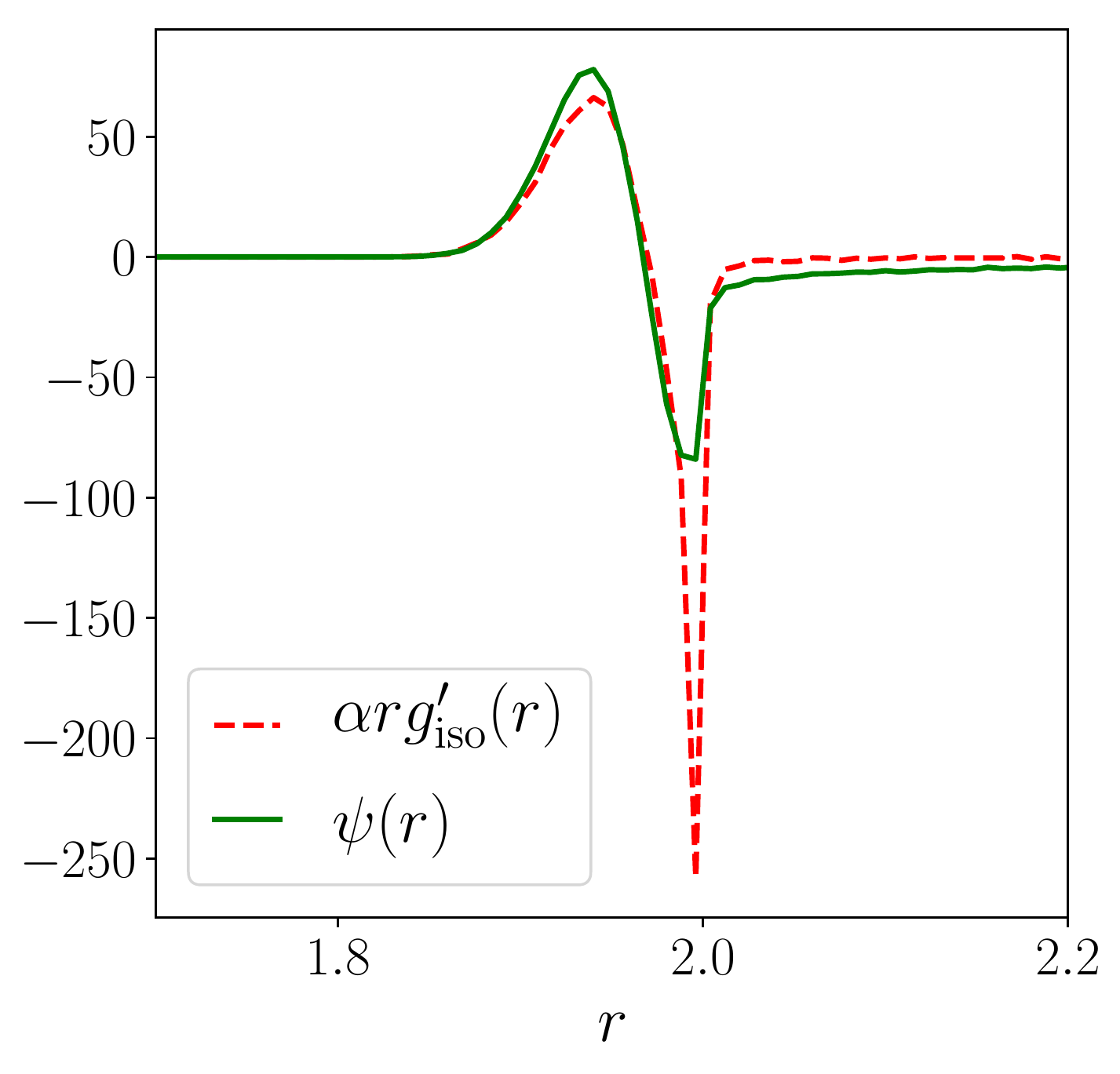}
		\caption{\comnc{(left)} Plot of the isotropic part $\giso(r)$ of the pair correlation function, measured in the stationary state of our numerical simulation of a sheared two-dimensional harmonic spheres ($\phi=0.875$, $\dot{\gamma}\tau_0=5\times10^{-4}$). Insets display the angular plots of $g(r,\theta)$ at fixed values of $r$ indicated by dots on the curve $\giso(r)$. The numerically determined $g(r,\theta)$ appears in full lines, while the parameterization Eq.~\eqref{DL_g} is indicated by dashed lines.
		The second order harmonics is seen to dominate the anisotropy, except for the value of $r$ which maximizes $\giso(r)$. \comnc{(right) Comparison between the functions $\psi(r)$ (Eq.~\eqref{DL_g_pre}) and $\alpha r \giso'(r)$ measured in our numerical simulations of a sheared two-dimensional suspension of harmonic disks ($\phi=0.875$, $\dot{\gamma}\tau_0=5\times10^{-4}$). }}
		\label{fig:gr}
	\end{center}
\end{figure}
This structure tensor can be conveniently used to perform a weakly anisotropic expansion of $g(\mathbf{r})$  close to the isotropic pair correlation function $\giso(r)$.
Considering $g(\mathbf{r})$ as a function of $r$ and $\er$, where the dependence on $\er$ characterizes the anisotropy, we expand $g(\mathbf{r})$ to quadratic order in $\er$
(corresponding to linear order in $\gQ$), leading to
\begin{equation}
	\label{DL_g_pre}
		g(\R{}) \approx \giso(r) + \psi(r) \left(\gQ:\eroer\right),
\end{equation}
where the $r$-dependence of the anisotropic term in the expansion is encoded into a function $\psi(r)$ to be determined.
An approximate expression of $\psi(r)$ can be found as follows.
Considering a small deformation, we assume that the
pair correlation function $g(\mathbf{r})=g(r,\theta)$ can be approximated by a direction-dependent, homothetic transformation of the isotropic pair correlation function $\giso(r)$, generated by the strain-rate tensor $\Einf$.
In particular, the first shell of neighbors deforms into an approximately elliptic shape, as sketched in Fig.~\ref{fig:ellipse}.
The position of the first-neighbor peak can be parameterized with the direction $\theta$ of the vector $\R{}$ as 
\begin{equation}
	\label{first_neighbour_peak_position}
	r_0^{}(\theta)=\rs\big(1-\alpha\left(\gQ:\eroer\right)\big),
\end{equation}
where $\rs$ is the angle-average position of the first-neighbor peak and $\alpha$ a proportionality factor to be determined self-consistently.
In the weakly anisotropic regime, we thus assume the pair correlation function along the direction $\theta$ to be only a homothetic transformation of the isotropic pair correlation function $\giso(r)$:
\begin{equation}
	\label{homothety_g}
	g(\R{})=\giso \left(\frac{r}{r_0(\theta)/\rs}\right).
\end{equation}
We also assume that the anisotropy is small enough so that the first shell of neighbors remains everywhere in contact with the focus particle (i.e., $r_0(\theta)<2$).
Expanding Eq.~(\ref{homothety_g}) to leading order in $\gQ$ then yields
\begin{equation}
	\label{DL_g}
	g(\R{}) \approx \giso(r) + \alpha r \giso'(r)\left(\gQ:\eroer\right)
\end{equation}
whence the explicit expression $\psi(r) = \alpha r \giso'(r)$ follows by comparison with Eq.~(\ref{DL_g_pre}).
Injecting the expression (\ref{DL_g}) of $g(\R{})$ into the definition (\ref{eq:def:Q}) of $\gQ$, a self-consistency condition fixes the value of $\alpha$ through the relation
\begin{equation} \label{eq:self-consistency}
\alpha \int_0^2 r^4 \giso'(r) \D r = \frac{4}{\pi\rho^2}\,.
\end{equation}
The pair correlation function measured in our numerical simulations (see Fig.~\ref{fig:gr}) shows that the second order angular harmonics already captures the leading anisotropy, except at the $r$ value where the isotropic pair correlation function $\giso(r)$ is maximum, in which case higher order modes become more visible (Fig.~\ref{fig:gr} left). In particular, it captures the swap of principal axes of the microstructure around the location of the maximum of $\giso(r)$, corresponding to a change of sign of $\psi(r)$ in Eq.~\eqref{DL_g_pre}.
\comnc{In Fig.~\ref{fig:gr} right, we explicitly compare the $r$-dependence of the anisotropic part of $g(\R{})$, measured by $\psi(r)$, to the parameterization \eqref{DL_g}. 
It shows that $\psi(r)$ is reasonably well approximated by $\alpha r \giso'(r)$, with no free parameters. Our approximation notably captures the characteristic oscillation of $\psi(r)$ localized around $r^\ast$. However, the minimum observed for $r\approx 2$ is overestimated by our approximation.}

\subsection{\label{subsec:integrals_g}Expression of tensors defined as integrals of \texorpdfstring{$g$}{g}}

Using the approximation~(\ref{DL_g}), we now express as a function of $\gSigma'$ all the tensors appearing in the r.h.s.~of Eq.~(\ref{eq_sigmaprim}) that are defined as integrals of $g$.
To proceed further, we will need the following properties, valid for any pair of symmetric traceless tensors $\gA$ and $\gB$:
\begin{align}
	\label{integralg_prop1}
	& \int_{-\pi}^{\pi}\left(\eroer - \frac{1}{2} \identity \right) \D\theta=\mathbf{0}, \\
	\label{integralg_prop2}
	& \int_{-\pi}^{\pi}\left(\gA:\er \otimes \er\right)\D\theta=0, \\
	\label{integralg_prop3}
	& \int_{-\pi}^{\pi}\left(\eroer - \frac{1}{2} \identity \right) \left(\gA:\er \otimes \er\right)\D\theta=\frac{\pi}{2}\gA, \\
	\label{integralg_prop4}
	& \int_{-\pi}^{\pi}\left(\gA:\er \otimes \er\right)\left(\gB:\er \otimes \er\right)\D\theta=\frac{\pi}{2}\left(\gA:\gB\right), \\
	\label{integralg_prop5}
	& \int_{-\pi}^{\pi}\!\left(\!\eroer \!- \frac{1}{2} \identity \right)\!\left(\gA\!:\!\er \otimes \er\right)\left(\gB\!:\!\er \otimes \er\right)\D\theta=\mathbf{0}.
\end{align}
Using these properties, we get a linear relation between the deviatoric part of stress tensor and the structure tensor, as well as an expression of the pressure:
\begin{align}
	\label{rel_sigmap_qp}
	& \gSigma'= \left(\frac{\alpha\pi\rho^2}{4}\int_0^2r^3f(r)g_{\mathrm{iso}}'(r)\D r\right) \gQ \equiv k \gQ, \\
	\label{rel_p_giso}
	& p= -\frac{\pi\rho^2}{2}\int_0^2r^2f(r)g_{\mathrm{iso}}(r)\D r.
\end{align}
We also get the following expressions for the tensors $\gTheta$, $\gPhi$, $\gXi$ and $\gPi$ defined in
Eqs.~(\ref{def_Theta}), (\ref{def_Phi}), (\ref{def_Xi}) and (\ref{def_Pi}),
\begin{align}
	\gTheta' & = \left(\frac{\pi\rho^2}{4}\int_0^2r^3f'(r)\giso(r)\D r\right) \Einf, \\
	\tr\gTheta & = \left(\frac{\pi\alpha\rho^2}{4k}\int_0^2r^4f'(r)\giso'(r)\D r\right) \left(\Einf : \gSigma'\right), \\
    \gPhi' & = \left(\frac{\pi\rho^2}{4}\int_0^2r^2f(r)\giso(r)\D r\right) \Einf = -\frac{p}{2}\,\Einf, \\
	\tr\gPhi & = \left(\frac{\pi\alpha\rho^2}{4k}\int_0^2r^3f(r)\giso'(r)\D r\right) \left(\Einf : \gSigma'\right), \\
	\gXi' & = \left(\frac{\pi\alpha\rho^2}{4k}\int_0^2r^2f^2(r)\giso'(r)\D r\right) \gSigma', \\
	\tr\gXi & = \pi\rho^2\int_0^2rf^2(r)\giso(r)\D r, \\
	\gPi' & = \left(\frac{\pi\alpha\rho^2}{4k}\int_0^2r^3f(r)f'(r)\giso'(r)\D r\right) \gSigma', \\
	\tr\gPi & = \pi\rho^2\int_0^2r^2f(r)f'(r)\giso(r)\D r.
\end{align}

\subsection{\label{subsec:integrals_g3}Kirkwood closure and expression of tensors defined as integrals of \texorpdfstring{$\gthree$}{g3}}

\begin{figure}
	\begin{center}
		\includegraphics[width=\linewidth]{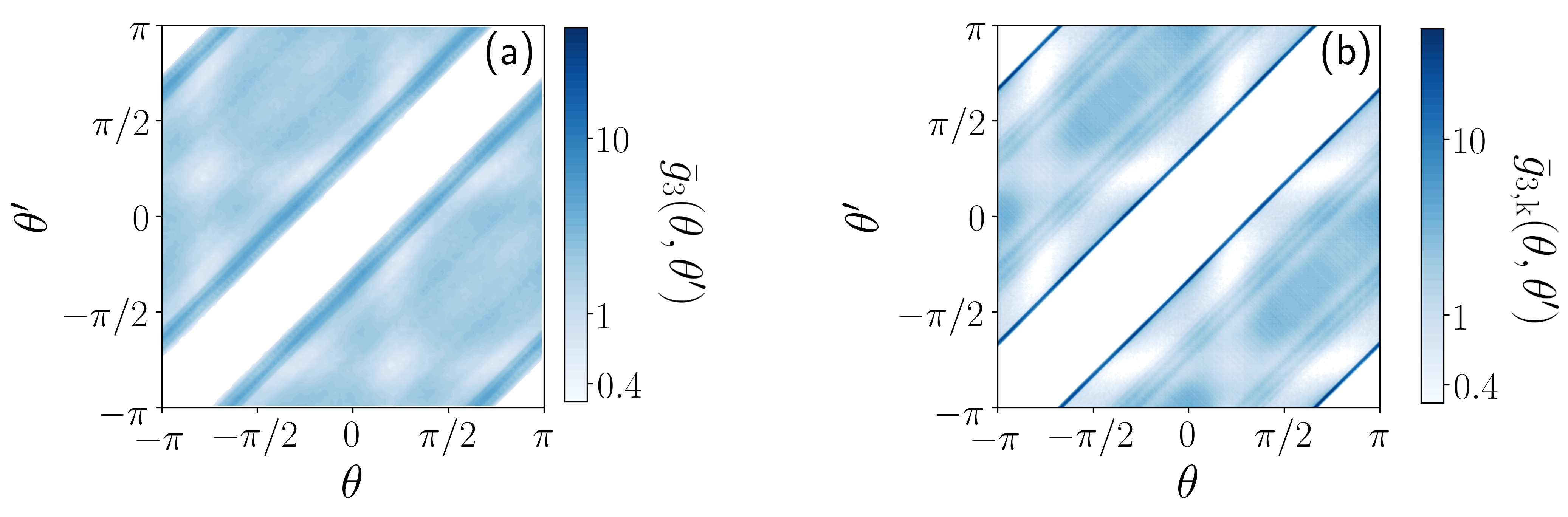}
		\includegraphics[width=0.5\linewidth]{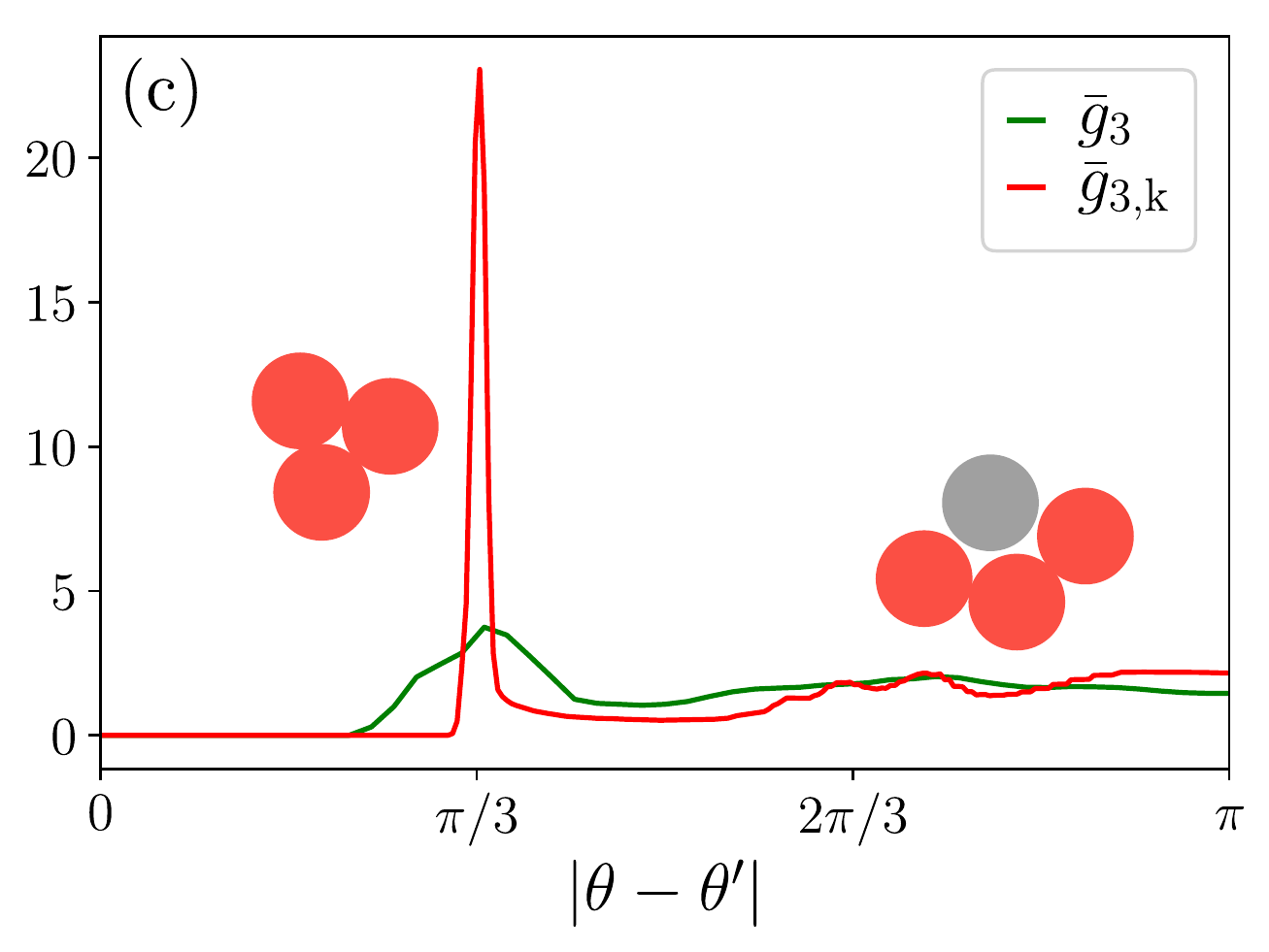}
		\caption{Test of the Kirkwood closure in numerical simulations of a suspension at $\phi=0.875$ sheared at $\dot{\gamma}\tau_0=5\times10^{-4}$ (see text for numerical details).
		(a) Colormap of $\bar{g}_3(\theta, \theta')$ (see Eq.~\ref{eq:contact_avg_g3} for definition). White areas correspond to $\bar{g}_3=0$.
		(b) Colormap of the Kirkwood approximation $\bar{g}_{3,\mathrm{K}}(\theta, \theta')$ (see Eq.~\ref{eq:contact_avg_g3K} for definition). (c) Evolution of $\bar{g}_3$ and its Kirkwood approximation $\bar{g}_{3,\mathrm{K}}$ as a function of $|\theta-\theta'|$ averaged over all the values of $\theta$.}
		\label{fig:kirkwood}
	\end{center}
\end{figure}

To evaluate the remaining tensors $\gGamma$ and $\gUpsilon$ defined in Eqs.~(\ref{def_Gamma}) and (\ref{def_Upsilon}), we need to make some assumptions on $\gthree$. We choose the well-known Kirkwood closure \cite{kirkwood_statistical_1935} that links $\gthree$ to $g$ as:

\begin{equation}
	\label{kirkwood}
	\gthree(\R{},\rprim)=g(\R{})g(\rprim)g(\R{}-\rprim) \,.
\end{equation}
This approximation is widely used in the physics of liquids and is known to be quite accurate for dilute systems, which is not the regime considered here. However, in practice we need the closure to be approximately valid only in situations where two particles are in contact with the same third particle. In other cases the force term is zero so that such cases do not contribute to the integrals defining $\gGamma$ and $\gUpsilon$. 

We test the Kirkwood closure for these contact situations in our simulations.
We define the average over contacts of $\gthree(\bm{r}, \bm{r}')$,
\begin{equation}
    \bar{g}_3(\theta, \theta') = \iint_{r<2, r'<2} r \mathrm{d} r \, r'\mathrm{d} r' \gthree(\bm{r}, \bm{r}'),
    \label{eq:contact_avg_g3}
\end{equation}
with $\theta=\arg(\bm{r})$ and $\theta'=\arg(\bm{r}')$, and of the Kirkwood closure,
\begin{equation}
    \bar{g}_{3,\mathrm{k}}(\theta, \theta') = \iint_{r<2, r'<2} r \mathrm{d} r \, r'\mathrm{d} r' g(\bm{r}) g(\bm{r}') g(\bm{r}-\bm{r}').
    \label{eq:contact_avg_g3K}
\end{equation}
We show in Fig.~\ref{fig:kirkwood} a comparison between the measured correlation function $\bar{g}_3(\theta, \theta')$ and its Kirkwood closure approximation $\bar{g}_{3,\mathrm{K}}(\theta, \theta')$, averaged over $300$ statistically independent configurations of a system at $\phi=0.875$
(with $\phi_{\rm J}$ the estimated jamming packing fraction)
sheared at $\dot{\gamma}\tau_0=5\times10^{-4}$.
The Kirkwood closure captures \reveb{reasonably} well the qualitative features of $\bar{g}_3$, such as local maxima and minima, as shown in Fig.~\ref{fig:kirkwood}a-b.
In~Fig.~\ref{fig:kirkwood}c, taking advantage of the quasi symmetry around the axis $\theta = \theta'$, we average $\bar{g}_3$ and $\bar{g}_{3,\mathrm{K}}$ over $\theta$, leaving functions depending only on $|\theta - \theta'|$.
This shows that quantitatively, the Kirkwood closure both overestimates the value of the peak of $\bar{g}_3$ around $|\theta - \theta'| = \pi/3$ (which corresponds to three particles in contact with one another, as sketched in Fig.~\ref{fig:kirkwood}c) and underestimates its width, so that its weight is reasonably well captured. 
\reveb{Fig.~\ref{fig:kirkwood}c also shows that for larger angular differences, $|\theta - \theta'| \apprge 2\pi/3$, the Kirkwood closure} gives a \reveb{fair} approximation of the probability of having three particles $i, j$ and $k$ such that $i$ is in contact with $j$ and $k$, but $j$ and $k$ are separated by at least another particle $l$.
\reveb{At any rate, we stress again that the first motivation to use the Kirkwood closure is that it is the simplest closure satisfying the required permutation symmetries. Indeed, using a more involved closure would lead to extremely complicated calculations to derive the constitutive model. The goal of the above numerical comparison was thus to check that the Kirkwood closure can be considered as a reasonable approximation at a qualitative level, and not to assess the approximation in a quantitative way.}

Replacing $g$ by its expression (\ref{DL_g}) in the closure relation (\ref{kirkwood}) and using Eq.~(\ref{rel_sigmap_qp}) to relate $\gSigma'$ and $Q$, we are now able to link $\gGamma$ and $\gUpsilon$ to $\gSigma'$ (details of the calculation can be found in \ref{app:calc_integrals}):

\begin{align}
	\label{expr_gammap}
	\gGamma' & =\Gamma_1 \gSigma' + \frac{\Gamma_3}{2} \left(\gSigma':\gSigma'\right)\gSigma', \\
	\label{expr_trgamma}
	\tr\gGamma & = 2\Gamma_0 + \Gamma_2 \left(\gSigma':\gSigma'\right), \\
	\label{expr_upsilonp}
	\gUpsilon' & =\Upsilon_1 \gSigma' + \frac{\Upsilon_3}{2} \left(\gSigma':\gSigma'\right)\gSigma', \\
	\label{expr_trupsilon}
	\tr\gUpsilon & = 2\Upsilon_0 + \Upsilon_2 \left(\gSigma':\gSigma'\right),
\end{align}
where $\Gamma_{i}$ and $\Upsilon_{i}$ are coefficients whose expressions are given in \ref{app:calc_integrals}.

Replacing the tensors $\gTheta$, $\gPhi$, $\gXi$, $\gPi$, $\gGamma$ and $\gUpsilon$
by their expressions in the evolution equations (\ref{eq_sigmaprim}) for $\gSigma'$ and (\ref{eq_p}) for $p$, we get:

\begin{align}
	\label{eq_sigmaprim_closed}
	&\dot{\gSigma}' = \kappa\{\giso\} \Einf + \Ominf\gdot\gSigma' - \gSigma'\gdot\Ominf + \left[\beta\{\giso\} - \xi\{\giso\} \left(\gSigma':\gSigma'\right)\right]\gSigma' \,, \nonumber\\
	\\
	\label{eq_p_closed}
	& \dot{p} = \zeta\{\giso\}\left(\Einf:\gSigma'\right) + \eta\{\giso\} + \chi\{\giso\} \left(\gSigma':\gSigma'\right),
\end{align}
where the coefficients $\kappa$, $\beta$, $\xi$, $\zeta$, $\eta$, and $\chi$ are functionals of $\giso$ (which may therefore have an implicit time dependence) defined as:

\begin{align}
	\label{def_A}
	& \kappa\{\giso\} = \frac{\pi\rho^2}{4}\int_0^2r^3f'(r)\giso(r)\D r - \frac{3p}{2}, \\
	\label{def_B}
	& \beta\{\giso\} =-\frac{\pi\alpha\rho^2}{4k}\int_0^2r^2\left(f^2(r)+rf(r)f'(r)\right)\giso'(r)\D r - \Gamma_1 - \Upsilon_1  \\
	\label{def_C}
	& \xi\{\giso\} = \frac{\Gamma_3}{2} + \frac{\Upsilon_3}{2}, \\
	\label{def_D}
	& \zeta\{\giso\} = \frac{\pi\alpha\rho^2}{8k}\int_0^2r^3\left(f(r)-rf'(r)\right)\giso'(r)\D r -1, \\
	\label{def_E}
	& \eta\{\giso\} =\frac{\pi\rho^2}{2}\int_0^2r\left(f^2(r)+rf(r)f'(r)\right)\giso(r)\D r + \Gamma_0 + \Upsilon_0, \\
	\label{def_F}
	& \chi\{\giso\} = \frac{\Gamma_2}{2} + \frac{\Upsilon_2}{2}
\end{align}

It is interesting to note that all the tensors and scalar invariants allowed by frame indifference and algebraic considerations \cite{hand_theory_1962} in two dimensions appear in our evolution equation for $\gSigma'$ and $p$.

\reveb{At this stage, we have found evolution equations for $\gSigma'$ and $p$, Eqs.~(\ref{eq_sigmaprim_closed}) and (\ref{eq_p_closed}), that depend on the microstructure through the isotropic pair correlation function $\giso$. Several routes can be followed to determine explicitly the values of the coefficients $\kappa$, $\beta$, $\xi$, $\zeta$, $\eta$, and $\chi$. In the following, we use a simple parametrization of $\giso$ to obtain approximate analytical expressions of these coefficients. Alternatively, one may also determine numerically $\giso$ in a molecular dynamics simulation in order to evaluate the coefficients appearing in Eqs.~(\ref{eq_sigmaprim_closed}) and (\ref{eq_p_closed}). Other approaches inspired from liquid theory might also be considered to determine approximations of $\giso$. These alternative routes could possibly be explored in future works.}

%\textcolor{red}{PARAGRAPH sur $g_\mathrm{iso}$ numérique}

\section{\label{sec:calculation_coefficient}Calculation of the coefficients}

\subsection{\label{subsec:parametrization_giso}Parametrization of the isotropic part of the pair correlation function}

To evaluate the coefficients $\kappa$, $\beta$,\dots $\chi$ defined in Eqs.~(\ref{def_A})--(\ref{def_F}), we need to know the shape of the isotropic part of the pair correlation function $\giso(r)$.
%One possibility would be to measure $\giso(r)$ in molecular dynamics simulations and to evaluate the coefficients  numerically.
To make progress on the analytical side, we use a simple parametrization of $\giso(r)$.
Drawing inspiration from the typical form of the pair correlation function in molecular dynamics simulations, we approximate $\giso(r)$ as a Dirac peak at position $\rs\le2$ plus a Heaviside function above $\rs$:

\begin{equation}
	\label{expr_giso}
	\giso(r)=\frac{A}{\rs}\delta(r-\rs)+H(r-\rs)
\end{equation}

\begin{figure}
	\begin{center}
		\includegraphics[width=0.49\columnwidth]{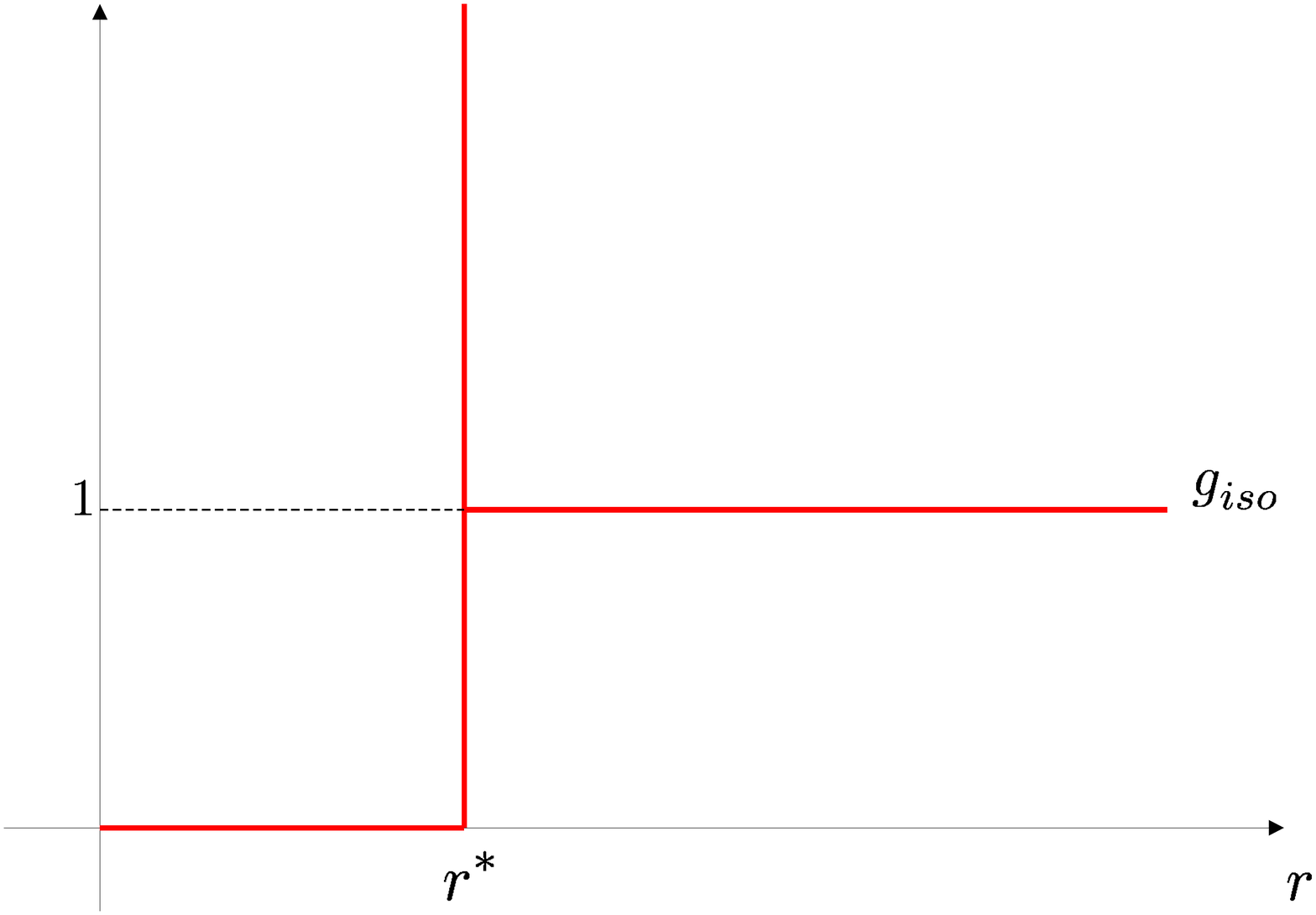}
		\includegraphics[width=0.49\columnwidth]{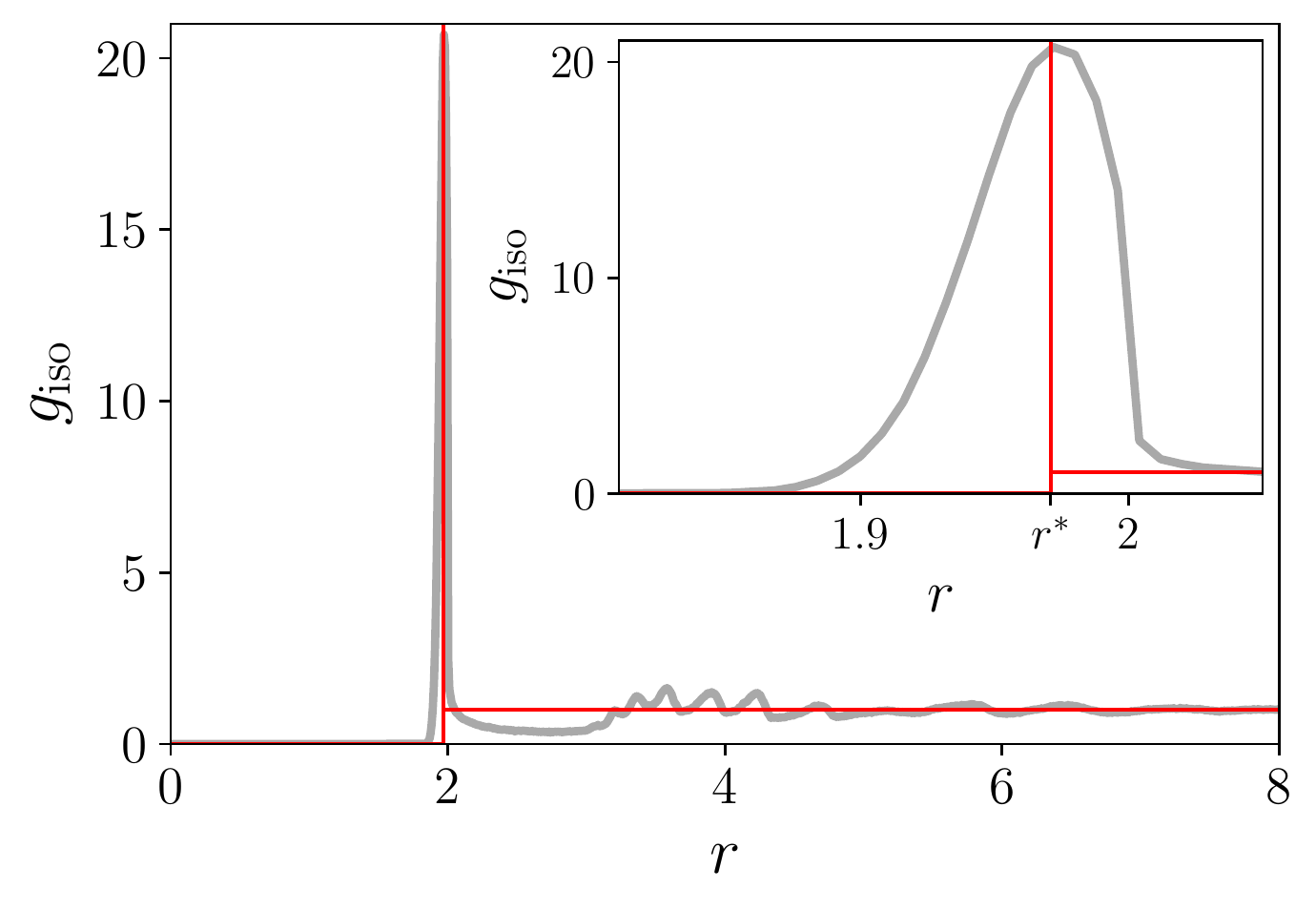}
		\caption{(left) Schematic representation of the parametrization of $\giso(r)$ given in Eq.~(\ref{expr_giso}). (right) Comparison of our parametrization of $\giso(r)$ (in red) with the isotropic part of the pair correlation function measured in \comnc{simulations of a suspension at $\phi=0.875$ sheared at $\gd\tau_0=5\times10^{-4}$} (in gray). Inset: zoom on the peak of $\giso(r)$.}
		\label{giso}
	\end{center}
\end{figure}

The Dirac peak models the first shell of neighbors observed in the pair correlation function, while the Heaviside function models the diffuse density background beyond the first shell, assumed to be structureless (i.e., the second shell of neighbors, whose amplitude is weak in numerical simulations, is not taken into account).
We assume there is no particle at a distance shorter than the first-neighbor peak, so $\giso(r)=0$ for $r<\rs$. 
The prefactor $A/\rs$ in front of the Dirac distribution accounts for a constant number of neighbor particles in the first shell, equal to $2\pi\rho A$.
We allow $\rs$, the position of the first-neighbor peak, to vary with time. Assuming the average number of particles in the first shell to be equal to 6~\cite{schreck_tuning_2011}, we set the value of A to

\begin{equation}
	\label{expr_A}
	A=\frac{3}{\pi\rho}.
\end{equation}
Under this parametrization of $\giso(r)$, the self-consistency relation \eqref{eq:self-consistency} constraining $\alpha$ yields the explicit expression

\begin{equation}
	\label{expr_lambda}
	\alpha = \frac{4}{\pi\rho^2 {\rs}^2\left({\rs}^2-4A\right)}
\end{equation}
Eqs.~\eqref{expr_giso} and \eqref{expr_lambda} allow us to express the coefficients $\kappa$, $\beta$, $\xi$, $\zeta$, $\eta$ and $\chi$ as functions of $\rs$, given an explicit expression of the repulsive force $f(r)$. As a simple representative example, we choose a linear (dimensionless) force $f(r)=r-2$ (however, our approach is also valid for a more general force). We then obtain the following expression of the pressure in terms of $\rs$,

\begin{equation}
	\label{expr_p_rs}
	p=\frac{\pi\rho^2}{24}\left(\rs-2\right)\left(3\rs^3-2\rs^2-4(3A+1)\rs-8\right)
\end{equation}

We also obtain explicit (but lengthy) expressions in terms of $\rs$ of the coefficients $\kappa$, $\beta$,\dots $\chi$ appearing in the evolution equations (\ref{eq_sigmaprim_closed}) and (\ref{eq_p_closed}) for $\gSigma'$ and $p$ respectively.
The full expressions of these coefficients are given in \ref{app:exp_coef}.

Different strategies can be followed at this stage. A first one is to use Eq.~\eqref{expr_p_rs} to transform the evolution equation (\ref{eq_p_closed}) on $p$ into an evolution equation on $\rs$, yielding two coupled evolution equations on $\gSigma'$ and $\rs$, which can be integrated numerically.
A second strategy is to keep the pressure $p$ as the relevant dynamical variable, and to expand Eq.~\eqref{expr_p_rs} to either first or second order in terms of the small parameter $2-\rs$, assuming that the pressure is small (i.e., the system is only slightly above jamming).
Details of these expansions and of the obtained expressions of the coefficients in terms of $p$ are given in \ref{app:exp_coef}.
The coupled evolution equations (\ref{eq_sigmaprim_closed}) and (\ref{eq_p_closed}) on $\gSigma'$ and $p$ then become closed, 
\begin{align}
	\label{eq_sigmaprim_closed_p}
	&\dot{\gSigma}' = \kappa(p) \Einf + \Ominf\gdot\gSigma' - \gSigma'\gdot\Ominf  + \left[\beta(p)- \xi(p) \left(\gSigma':\gSigma'\right)\right]\gSigma' \,, \\
	\label{eq_p_closed_p}
	& \dot{p} = \zeta(p) \left(\Einf:\gSigma'\right) + \eta(p)+ \chi(p) \left(\gSigma':\gSigma'\right),
\end{align}
and can be integrated numerically \reveb{(see below and \ref{app:evol_p} for brief discussions of the behavior of these coupled equations)}.
In the next section, we further simplify these equations to reduce the description to a single equation on $\gSigma'$.

\subsection{\label{subsec:reduction_tensorial_equation}Reduction to a single tensorial equation}

To simplify the description, one may take advantage of the time scale separation between the dynamics of $p$ and that of $\gSigma'$.
It is possible to show (see, e.g., the polar decomposition of $\gSigma'$ given in \cite{CunyPRL21}) that for low shear rate $\dot\gamma \tau_0 \ll 1$ the alignment dynamics of $\gSigma'$ onto the strain-rate tensor $\Einf$ is slow. 
This is just a consequence of the fast relaxation of unbalanced elastic forces 
compared to the time scale of flow.
To be more specific, the rotation of $\gSigma'$ is slow, but the dynamics of $\gSigma':\gSigma'$
remains fast at low shear rates.
In addition, the dynamics of the pressure $p$ also remains fast for low shear rates as compared to the orientational dynamics of $\gSigma'$.
For the sake of simplicity, we neglect the contributions from the strain-rate tensor (assumed to be small) in the evaluation of the stationary value of the pressure (or of $\rs$). We thus aim at an equation of state $p(\phi)$, where $\phi$ is the packing fraction.
Under this assumption, evaluating $\gSigma':\gSigma'$ from the stationary solution of Eq.~\eqref{eq_sigmaprim_closed} leads to
\begin{equation}
	\label{stationary_r}
	\gSigma':\gSigma'=\frac{\beta(p)}{\xi(p)},
\end{equation}
if $\beta,\xi > 0$ (otherwise $\gSigma':\gSigma'= 0$ in the stationary state).
%Here, we have kept the expressions of the coefficients in terms of $\rs$, but they may be turned into functions of $p$ through the use of Eq.~\eqref{expr_p_rs}, either numerically or using an expansion of $p$ in powers of $2-\rs$.
%
Injecting Eq.~\eqref{stationary_r} in the stationary equation (\ref{eq_p_closed_p}) for the pressure, we get

\begin{equation}
	\label{stationnary_p}
	\eta(p) + \frac{\beta(p)\chi(p)}{\xi(p)} = 0.
\end{equation}

Although the solution of this equation cannot be obtained in closed analytical form, it is possible to obtain an analytical approximation of the stationary value of $p$ close to jamming. Indeed close to jamming the pressure is low so that we can use the expansion of coefficients $\beta$, $\xi$, $\eta$ and $\chi$ up to linear order in $p$ (see \ref{app:exp_coef}) to linearize Eq.~(\ref{stationnary_p}) and solve it for $p$.
This gives us a relatively complex expression for $p$ as function of $\phi$ in stationary state.
%which can be consulted in \ref{app:exp_coef}.
According to this expression $p$ starts to be positive for $\phi>\phi_{\rm J}=5/4$, where $\phi_{\rm J}$ is thus interpreted as the jamming packing fraction. Note that due to the approximations made, the resulting value $\phi_{\rm J}=1.25$ is approximately $50$\% greater than the expected value for a two-dimensional packing. 
To remain consistent with our expansion to leading order in $p$, we expand the obtained $p(\phi)$ to first order in $\phi-\phi_{\rm J}$,
\begin{equation}
	\label{relation_p_phi}
	p\approx 0.63\times(\phi-\phi_{\rm J}).
\end{equation}
The full expression of the prefactor is given in \ref{app:exp_coef}.
In this simplified setup for harmonic spheres, we end up with a single tensorial equation, the evolution equation \eqref{eq_sigmaprim_closed} for $\gSigma'$, 

\begin{equation}
	\label{eq_sigmaprim_single}
	\dot{\gSigma}' = \bar{\kappa}(\phi) \Einf  + \Ominf \gdot\gSigma' - \gSigma' \gdot\Ominf + \left[\bar{\beta}(\phi)- \bar{\xi}(\phi) \left(\gSigma':\gSigma'\right)\right]\gSigma' ,
\end{equation}
where $\bar{\kappa}(\phi)$, $\bar{\beta}(\phi)$ and $\bar{\xi}(\phi)$ are the value of coefficients $\kappa$, $\beta$ and $\xi$ evaluated in $p(\phi)$ according to the relation (\ref{relation_p_phi}). Expanding these coefficients to first order in $\phi-\phi_{\rm J}$ (for $\phi>\phi_{\rm J}$), we get the explicit numerical expressions:

\begin{align}
    \label{expr_bar_kappa}
    &\bar{\kappa}\approx 1.19 - 0.099\times (\phi-\phi_{\rm J})\,, \\
    \label{expr_bar_beta}
    &\bar{\beta}\approx 0.16 +0.76\times(\phi-\phi_{\rm J})\,, \\
    \label{expr_bar_xi}
    &\bar{\xi}\approx 0.62 +0.0054\times(\phi-\phi_{\rm J})\,.
\end{align}

\begin{figure}
    \centering
    \includegraphics[width=0.49\linewidth]{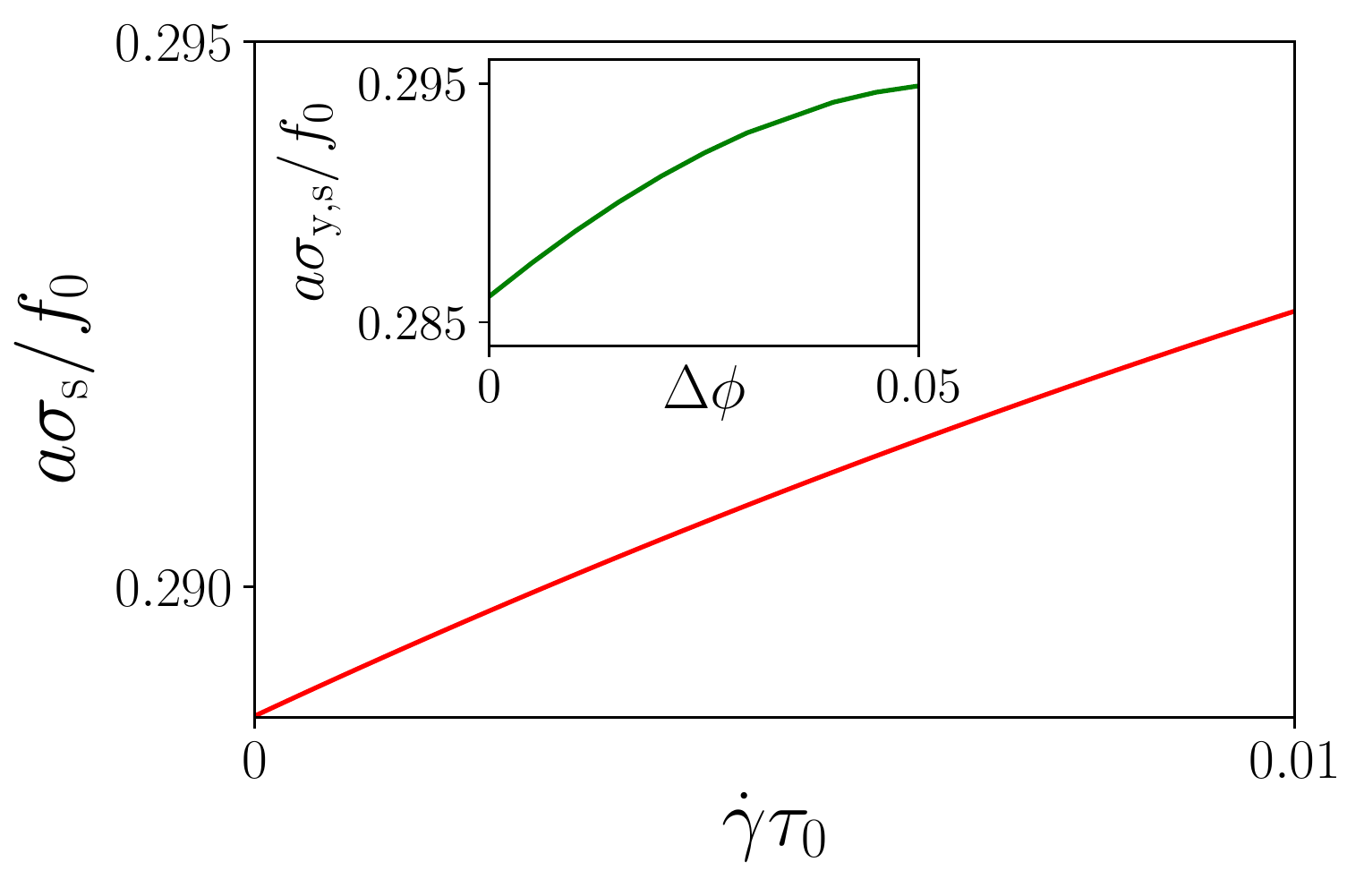}
    \includegraphics[width=0.49\linewidth]{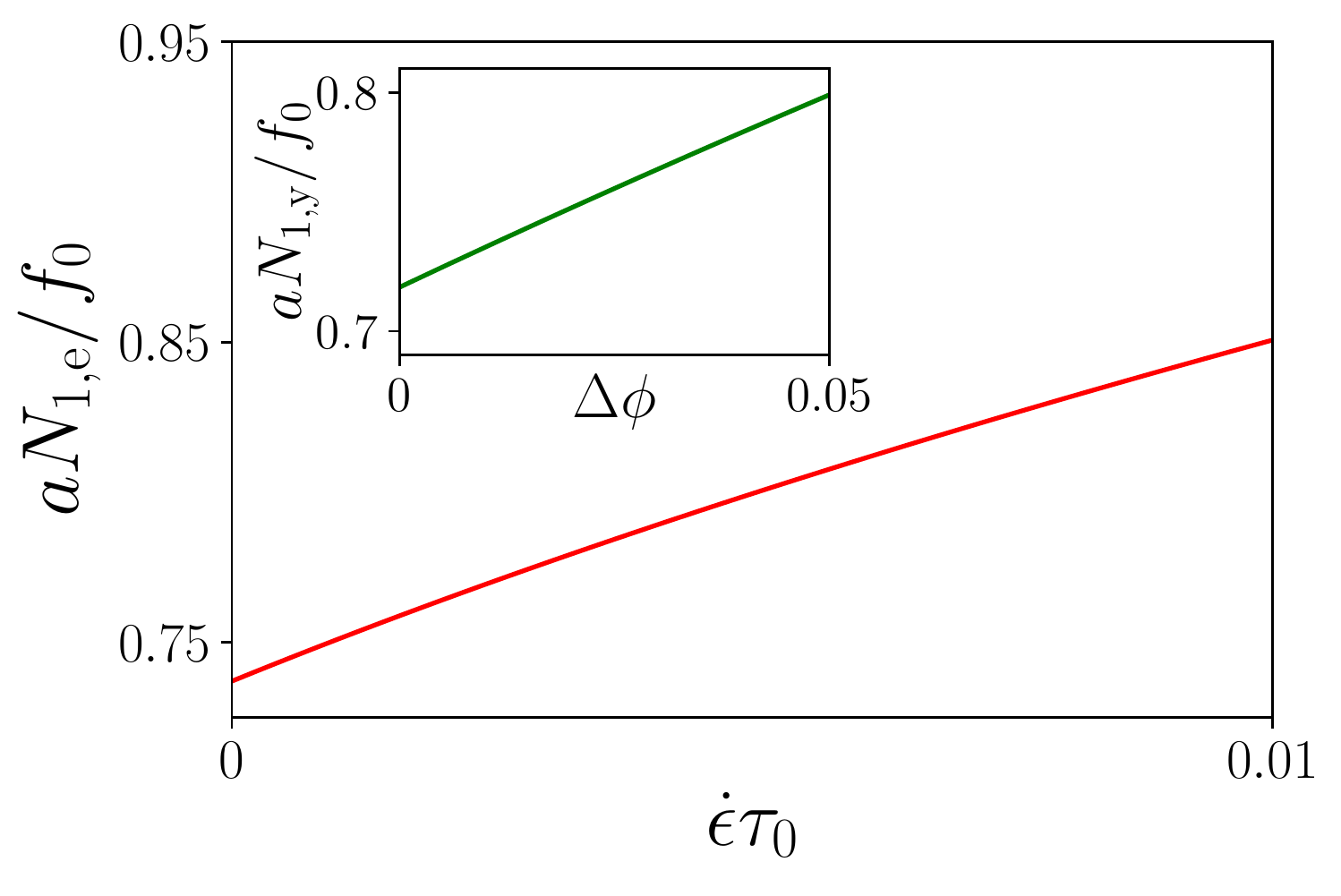}
    \caption{\comnc{(left) Predicted flow curve  in simple shear (shear stress $\sigma_\mathrm{s}$ versus shear rate $\dot\gamma\tau_0$) for $\Delta\phi=0.01$. Inset: yield stress $\sigma_\mathrm{s,y}$ as a function of the distance to jamming $\Delta \phi$. (right) Predicted flow curve for extensional flow (first normal stress difference $N_{1,\mathrm{e}}$ versus extensional rate $\dot\epsilon\tau_0$) for $\Delta\phi=0.01$. Inset: yield stress $N_{1,\mathrm{y}}$ as a function of the distance to jamming.}}
    \label{fig:flow_curves}
\end{figure}

%
%where the coefficients $\bar{\kappa}$, $\bar{\beta}$ and $\bar{\xi}$ are constants determined by the stationary value of $\rs$ as a function of $\phi$, in the low shear rate limit.
The already rich behavior of the evolution equation \eqref{eq_sigmaprim_single} on $\gSigma'$ is briefly discussed in \cite{CunyPRL21}, and detailed further in \cite{CunySM21}.
\reveb{The salient features of the steady rheology are summarized in Fig.~\ref{fig:flow_curves}, where we show the flow curves obtained under 
two different shear protocols.
In the left panel, we consider a simple shear, $(\nabla \bm{u}^\infty)_{ij} = \dot\gamma \delta_{ix}\delta_{jy}$, and plot the steady-state shear stress  
$\sigma_\mathrm{s} = \Sigma_{xy}$ as a function of the dimensionless shear rate $\dot\gamma\tau_0$.
It shows a typical yield stress behavior in both cases, with a finite value $\sigma_\mathrm{y,s}$ for the shear stress when $\dot\gamma \to 0$.
The behavior at small rates is the one of a Bingham fluid, that is, $\sigma_\mathrm{s} - \sigma_\mathrm{y,s} \propto \dot\gamma$, 
but at large rates the flow curves is concave.
This may be reminiscent of some experimental observations, however usually experiments support a Herschel-Bulkley behavior $\sigma_\mathrm{s} - \sigma_\mathrm{y,s} \propto \dot\gamma^n$, with $n<1$.
In inset we show the increase of the yield stress with the volume fraction.
Similarly, in the right panel we show a planar extensional flow, 
$(\nabla \bm{u}^\infty)_{ij} = \dot\epsilon (\delta_{ix}\delta_{jx} - \delta_{iy}\delta_{jy})$.
The flow curve is then the normal stress difference $N_{1,\mathrm{e}} = \Sigma_{xx} - \Sigma_{yy}$ as a function of the dimensionless extensional rate $\dot\epsilon \tau_0$.
It shows the same features as in simple shear, a Bingham-like behavior with a concave flow curve for large rates, and a yield normal stress difference $N_{1,\mathrm{y}}$ increasing with the volume fraction (inset).}

For completeness, we also investigated numerically the behavior of the coupled equations \eqref{eq_sigmaprim_closed_p} and \eqref{eq_p_closed_p} on $\gSigma'$ and $p$ respectively \reveb{(see \ref{app:evol_p})}. It turns out that these equations \reveb{unexpectedly predict that both the pressure $p$ and
the normal stress difference $N_1$
decay with the shear rate $\dot{\gamma}$.}
\reveb{A tentative physical interpretation of this result is given in \ref{app:evol_p}.}
\reveb{As a last comment, we note on general grounds} that including additional terms (typically higher order terms) in the continuum description of a system of interacting particles has also been reported in other contexts (e.g., thermal gases \cite{garcia-colinNavierStokesEquations2008} or systems of active particles \cite{peshkovBoltzmann2014}) to lead to unsatisfactory behaviors.

%%%%%%%%%%%%%%%%%%%%%%%%%%%%%%%%%%%%%%%%%%%%%%%%%%%%%%%%%%%%%%%%%%%%%%%%%%%%%%%%%%%%%%%%%%%%%%%%%%%%%%%%%%%%%%%%%%%%%%%%%%%%%%%%%%%%%%%%%%%%%%%%%%%%%%%%%%%%%%%%%%%%%%%%%%%%%%%%%%%%%%%%%%%%%%%%%%%%%%%%%%%%%%%%%%

\section{Effect of a small thermal noise}

We now try to go slightly beyond the purely athermal limit, and include a small thermal noise in the particle dynamics. We assume that the amplitude of the thermal noise remains small with respect to elastic forces, so that the system remains in an almost athermal regime.
\reveb{We use below the same approximation scheme as in the athermal case, which limits the validity of our approach to very small values of temperature.}

\reveb{We assume that the equation of motion of the particles in the thermal case are simply obtained by adding a Gaussian white noise term, as usually done in numerical simulations of dense suspensions of soft particles \cite{ikedaUnifiedStudyGlass2012,olssonAthermalJammingThermalized2013,kawasakiThinningThickeningMultiple2014}.}
The dimensionless equation for the particle dynamics now takes the form
\begin{equation}
	\label{dynamics:temp}
	-2 (\dot{\bm{r}}_\mu - \bm{u}^{\infty}_{\mu})+\bm{f}_{\mu} + \bm{\eta}_{\mu}(t)= \bm{0}\,,
\end{equation}
where $\bm{\eta}_{\mu}(t)$ is a Gaussian white noise satisfying
\begin{equation}
    \langle \eta_{\mu,i}(t)\eta_{\nu,j}(t')\rangle = 4 T \delta_{ij} \, \delta_{\mu\nu} \, \delta(t-t')\,.
\end{equation}
Here, the unit of temperature is $T_0=\frac{a f_0}{k_B}$.
\reveb{Note that the white noise term naturally comes out of the fluctuation-dissipation relation as soon as one assumes a simple friction term without memory. Conversely, the use of colored noise would imply a memory kernel for the friction term, leading to a much greater complexity of the calculations. Although strictly speaking the fluctuation-dissipation relation is valid only in the limit of vanishing shear rate, it is reasonable to neglect shear-rate-dependent corrections to the noise, again in line with standard numerical simulations of dense suspensions \cite{ikedaUnifiedStudyGlass2012,olssonAthermalJammingThermalized2013,kawasakiThinningThickeningMultiple2014}.}

\begin{figure}[t]
	\includegraphics[width=0.49\textwidth]{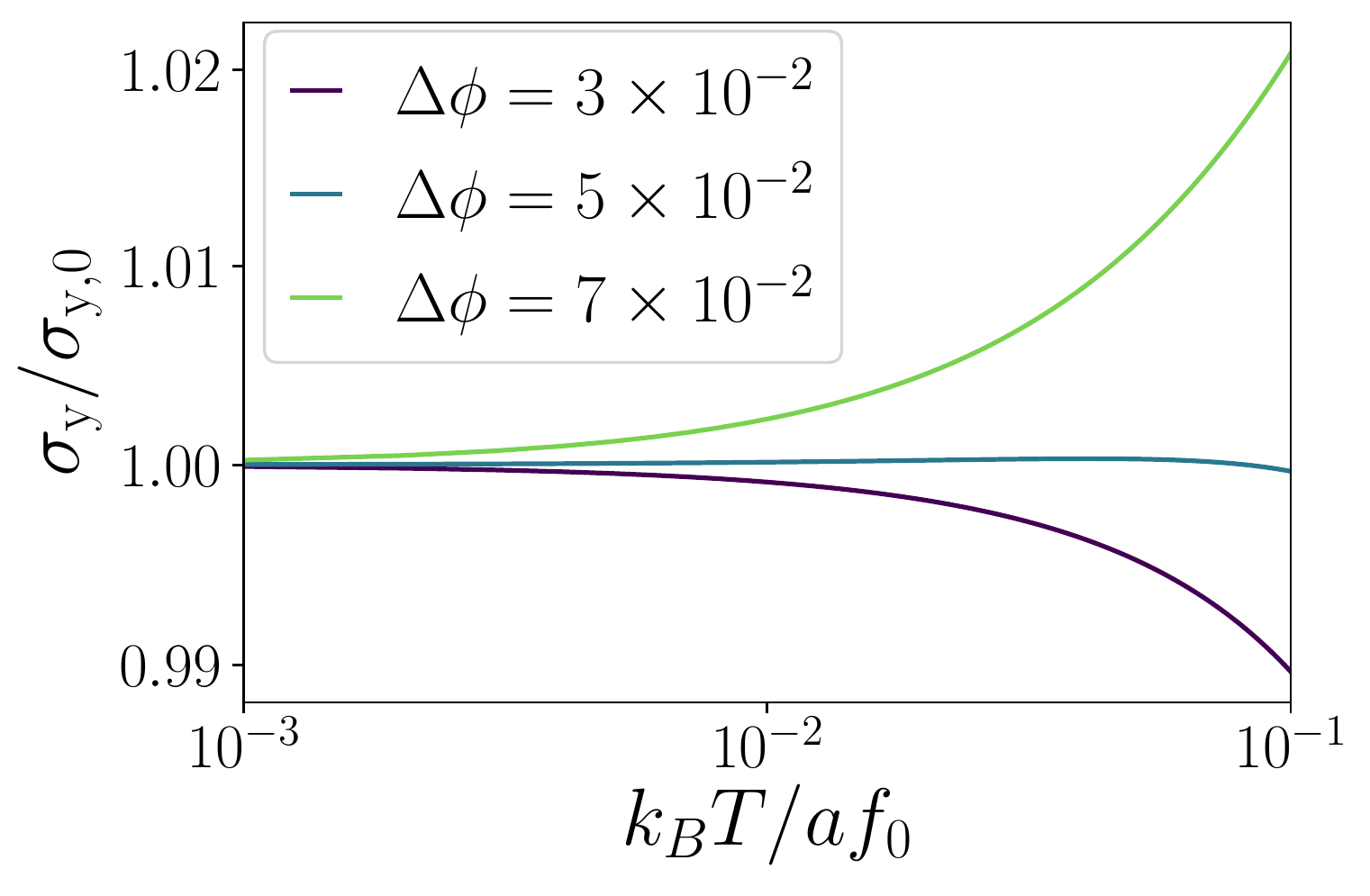}
	\hfill
	\includegraphics[width=0.49\textwidth]{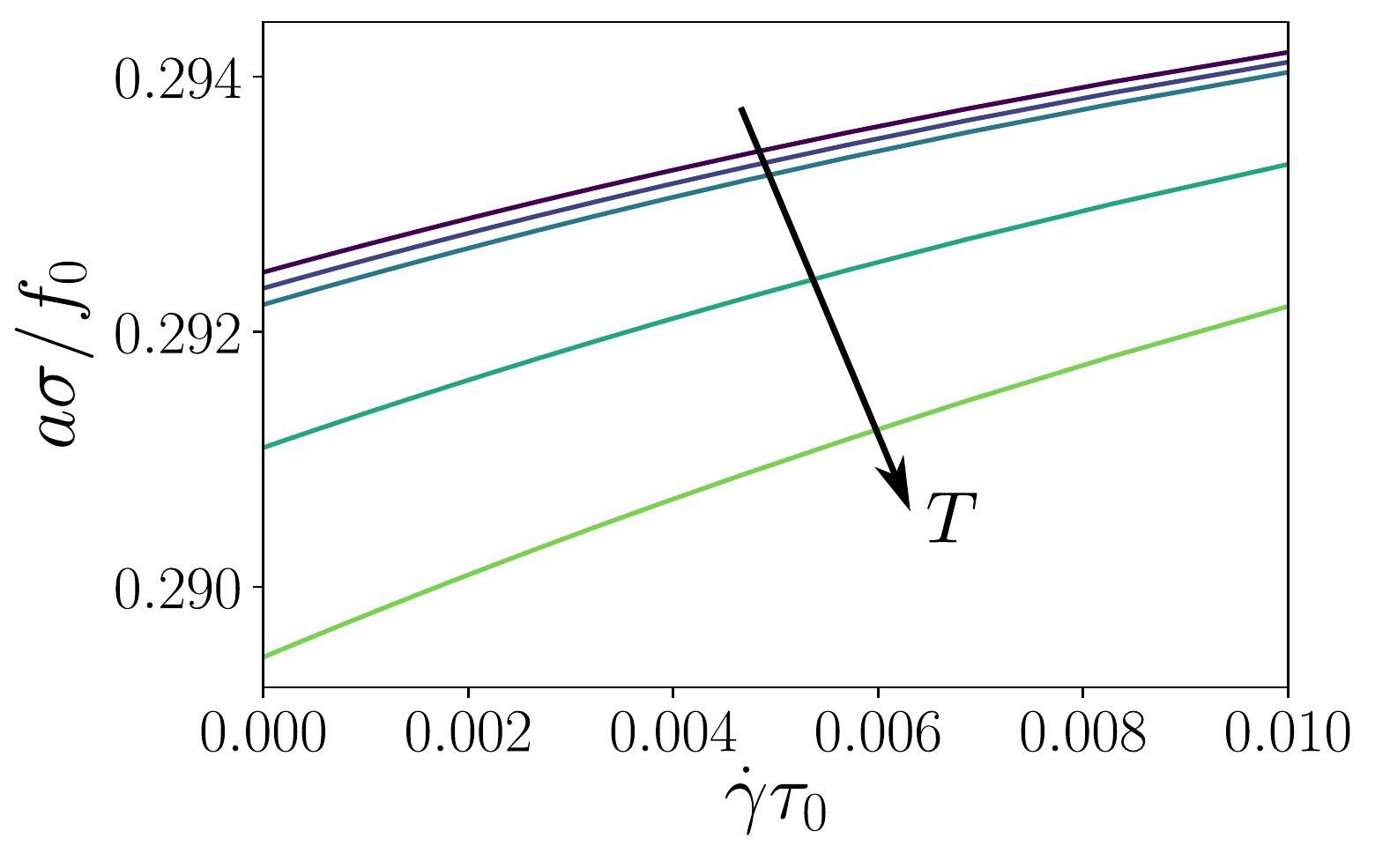}
	\caption{(Left) Yield stress $\sigma_y$ as a function of temperature for different $\Delta \phi$ ($\sigma_{y,0}$ is the zero-temperature yield stress value). (Right) Stationary flow curves for the shear stress obtained for $\Delta\phi=0.03$, displayed for different small values of temperature: $T=0,\,5.10^{-3},\,10^{-2},\,5.10^{-2},\,10^{-1}$.}
	\label{fig:stress:temp}
\end{figure}

The evolution equation for the stress tensor then acquires an additional term $\gPsi$ proportional to the temperature $T$ (detailed calculations are provided in \ref{app:temp})
\begin{equation}
	\label{eq_sigma:temp}
	\dot{\gSigma} = \DUinf \gdot\, \gSigma + \gSigma \gdot\, {\DUinf}^T + \mathbf{\Theta} - \mathbf{\Phi} - \mathbf{\Xi} - \mathbf{\Pi} - \mathbf{\Gamma} - \mathbf{\Upsilon} + \mathbf{\Psi},
\end{equation}
where $\gPsi$ is defined as
\begin{equation}
\gPsi=\frac{\rho^2 T}{2}\int\left(\rof\right)\nabla^2 g(\R{})\D\R{}.
\end{equation}
Using \reveb{the weakly anisotropic parametrization (\ref{DL_g}) of the correlation function $g(\R{})$, as well as the parametrization (\ref{expr_giso}) of the isotropic pair correlation function $g_{\rm iso}(r)$,} one can obtain explicit expressions of $\gPsi$ as a function of $\rs$, or of $p$ (see Eqs~\eqref{exp_Psi_Sigma_rstar:app:temp} and \eqref{exp_Psi_p:app:temp} in \ref{app:temp}).
Consistently with the approach developed in Sect.~\ref{subsec:reduction_tensorial_equation}, we neglect the effect of temperature on the pressure $p$, and keep the athermal equation of state $p(\phi)$.
The effect of temperature then boils down to an additional contribution to the coefficient $\beta$ in Eq.~\eqref{eq_sigmaprim_closed_p},
\begin{equation}
    \beta(T)=\beta(0) + 3T\left(\frac{A-4}{4A}+\frac{(A-2)(3A-4)}{4\pi A^3\rho^2}p\right).
\end{equation}
Depending on the packing fraction $\phi$, the resulting yield stress $\sigma_y$ either increases or decreases with temperature (see Fig.~\ref{fig:stress:temp}).
Stationary flow curves are also displayed on Fig.~\ref{fig:stress:temp} for different temperature values, for a packing fraction $\phi$ at which the stress decreases with temperature (fluidization effect).
We emphasize again that we only consider here the leading effect of a small temperature within an otherwise athermal physical picture, where elastic forces dominate over thermal fluctuations.

%%%%%%%%%%%%%%%%%%%%%%%%%%%%%%%%%%

\section{\label{sec:discussion}Discussion and conclusion}

To summarize our results, we have derived a nonlinear tensorial constitutive model from the particle-level dynamics of a two-dimensional dense soft suspension above jamming.
The obtained constitutive model takes the form of coupled nonlinear equations for the pressure and the deviatoric part of the stress tensor. A simplified version of the constitutive model has also been obtained in terms of a single tensorial equation, by taking advantage of a time scale separation which allowed us to get an equation of state for the pressure as a function of the packing fraction. The coefficients appearing in these equations have explicit expressions in terms of microscopic parameters and pressure.
This direct relation between the macroscopic description and the microscopic structure may help shedding light on the microscopic mechanisms at play (e.g., the shape of the soft repulsive force) in the macroscopic rheology. This issue \reveb{is} discussed in more details in a \reveb{companion paper} \cite{CunySM21}.

As discussed in \cite{CunyPRL21}, the resulting constitutive model exhibits even in its simplified version (with a single tensorial equation) a rich phenomenology typical of jammed soft suspensions, including (i) the existence of yield stresses for the shear stress as well as for the normal stress difference, (ii) an overshoot on the stress-strain curve upon a step change of shear rate, and (iii) a non-trivial dependence on the preshear rate of the residual stresses during relaxation after switching off the preshear, in qualitative agreement with experimental results
\cite{mohan_microscopic_2013,mohan_build-up_2014}.

The present work, mostly of methodological nature, potentially opens a new avenue in the study of dense suspensions close to jamming. While we have studied here the case of jammed soft suspensions, the methodology can be adapted in a relatively straightforward way to dense soft suspensions just below jamming.
Work in this direction is underway.
In addition, extending the present approach to the three-dimensional case would also be of great interest, and may lead to tensorial equations with additional terms as allowed by symmetry in three dimensions.
\reveb{This three-dimensional extension is expected to be relatively straightforward in its principles, as all the key steps and approximations can be translated to three-dimensional flows, or at least to pseudo-3D case where the flow is assumed to be invariant along the vorticity direction.}
If the presence of additional terms is confirmed, their effect on the rheological behavior will be worth investigating. Anyway, having a three-dimensional constitutive model at hand would be relevant for the comparison to experiments, even though we only aim at a qualitative agreement.

To improve the approach further, a desirable goal would be to obtain a Hershel-Bulkley behaviour as reported in most experiments, while the present constitutive model \reveb{essentially} describes a Bingham fluid \reveb{at low shear rate}, with an affine dependence on the shear rate. \reveb{Although on a broader range of shear rate the flow curve departs from a pure Bingham fluid and exhibits a small concavity, it remains far from the Hershel-Bulkley law (corresponding to constitutive law $\sigma= \sigma_y + a\dot{\gamma}^{1/2}$ for the shear stress component) that is widely used to describe experimental data.}
The reason for this discrepancy certainly lies in the important approximations made in the derivation of the constitutive model.
In particular, a crucial step in the derivation involves the closure of the three-body correlation function in terms of the pair correlation function. We have used here the basic Kirkwood relation as the simplest closure fulfilling the required symmetries of the three-body correlation function. We have checked numerically that the Kirkwood closure leads to a qualitatively correct angular structure of the three-body correlation function for two particles close to contact with a third one. Yet, the closure could certainly be improved at a quantitative level, at the price of more complicated calculations. For instance, one may wonder if including in the closure relation additional terms involving the derivative of the pair correlation function could be relevant.
In any case, a more sophisticated closure relation than the Kirkwood one may be required to account for a HB rheology.

Another key step in the derivation is the parametrization of the pair correlation function by the structure tensor, and the relation of the latter to the deviatoric part of the stress tensor. These different steps could in principle be improved in a somewhat systematic way, but again at the price of a significant additional complexity in the derivation. Finally, a more refined parametrization of the isotropic pair correlation function could also be used.
However, we note that although we took at each step the simplest option, with the aim to obtain a minimal constitutive model, the derivation presented here is already quite technical and lengthy.

Last but not least, another possible extension of our work could be to take into account a space dependence of the stress tensor and of the velocity gradient. Space heterogeneity may be important to describe, e.g., rotational flows or the shear-banding phenomenon.
It could be taken into account by assuming a scale separation between a microscopic length scale characterizing the typical distance between neighboring particles, and a macroscopic length scale characterizing the spatial variation of stresses and fluid flows.

%%%%%%%%%%%%%%%%%%%%%%%%%%%%%%%%%%%%%%%%%%%%%%%%%%%%%%%%%%%%%%%%%%%%%%%%%%%%%%%%%%%%%%%%%%%%%%%%%%%%%%%%%%%%%%%%%%%%%%%%%%%%%%%%%%%%%%%%%%%%%%%%%%%%%%%%%%%%%%%%%%%%%%%%%%%%%%%%%%%%%%%%%%%%%%%%%%%%%%%%%%%%%%%%%%

\section*{Acknowledgments}
This work is supported by the French National Research Agency in the framework of the "Investissements d'avenir" program (ANR-15-IDEX-02).

%%%%%%%%%%%%%%%%%%%%%%%%%%%%%%%%%%%%%%%%%%%%%%%%%%%%%%%%%%%%%%%%%%%%%%%%%%%%%%%%%%%%%%%%%%%%%%%%%%%%%%%%%%%%%%%%%%%%%%%%%%%%%%%%%%%%%%%%%%%%%%%%%%%%%%%%%%%%%%%%%%%%%%%%%%%%%%%%%%%%%%%%%%%%%%%%%%%%%%%%%%%%%%%%%%

\appendix

\section{\label{app:calc_lambda}Evaluation of tensorial terms}

In this Appendix, we provide further details on the evaluation of tensorial terms considered in Sec.~\ref{sec:eq_sigma}.

\subsection{Expression of the tensor \texorpdfstring{$\gLambda$}{Lambda}}

We start by evaluating the tensor $\gLambda$ introduced in Eq.~\eqref{def_Lambda}.
Expressing the gradient operator in polar coordinates $(r,\theta)$, we have for the radial force $\gF=f(r) \er$

\begin{equation}
	\label{def_DF}
		\nabla\gF = f'(r) (\er\otimes\er) + \frac{f(r)}{r} (\et\otimes\et)  = \left(f'(r)-\frac{f(r)}{r}\right) (\eroer) + \frac{f(r)}{r} \identity\,.
\end{equation}
We thus obtain:
\begin{align}
	\label{dvpt_lambda_integrande}
	(\ror)\gdot \big(\nabla \gF \gdot \nabla \Uinf \big)^T & = \frac{f(r)}{r} (\ror)\gdot\nabla\UinfT \nonumber\\
	& \qquad \qquad + \left(f'(r)-\frac{f(r)}{r}\right)(\ror) \gdot \big( (\eroer) \gdot \nabla \Uinf \big)^T \nonumber\\
	& = (\rof)\gdot\nabla\UinfT + r^2f'(r)(\eroer)\gdot \big( (\eroer) \gdot \nabla \Uinf \big)^T \nonumber\\
	& \qquad \qquad - rf(r) (\eroer) \gdot \big( (\eroer) \gdot \nabla \Uinf \big)^T 
\end{align}
It is then convenient to perform the following transformations,

\begin{align}
	\label{tensor_integrand}
	\left(\eroer\right) \gdot \big( (\eroer) \gdot \nabla \Uinf \big)^T &= \left(\eroer\right)\gdot\left(\er\otimes\left({\DUinf}^T\gdot\,\er\right)\right)^T \nonumber\\
	& = \left(\eroer\right)\gdot\left(\left({\DUinf}^T\gdot\,\er\right)\otimes\er\right) \nonumber\\
	& = \left(\er\gdot{\DUinf}^T\gdot\,\er\right)\left(\eroer\right)
\end{align}
as well as,

\begin{equation}
	\label{einf_eroer}
	\er\gdot{\DUinf}^T\gdot\,\er = {\DUinf}^T : \eroer = \Einf : \eroer  - \Ominf : \eroer = \Einf : \eroer.
\end{equation}
Using these results in the expression (\ref{def_Lambda}) of $\gLambda$, we finally get
Eq.~(\ref{dvpt_lambda_tens}).

\subsection{Decomposition of \texorpdfstring{$\DUinf \gdot\, \gSigma + \gSigma \gdot\, {\DUinf}^T$}{}}

Recalling that $\gSigma$ is a symmetric tensor, the trace of $\DUinf \gdot\, \gSigma + \gSigma \gdot\, {\DUinf}^T$ can be written as:

\begin{equation}
	\label{dem_tr1_step1}
	\tr \! \left(\DUinf \! \gdot \gSigma + \gSigma  \gdot \! {\DUinf}^T\right) = 2 \DUinf : \gSigma  = 2 \Einf : \gSigma  + 2 \Ominf : \gSigma
\end{equation}
Using the fact that $\Ominf$ is antisymmetric, it is easy to show that
$\Ominf : \gSigma$ is zero. In addition, we have:

\begin{equation}
	\label{dem_tr1_step2}
	\Einf : \gSigma = \Einf : \gSigma' + \frac{1}{2} \left(\tr\gSigma\right) \left( \tr\Einf \right) = \Einf : \gSigma' \,,
\end{equation}
as $\tr\Einf=\tr\DUinf=0$ since the fluid is incompressible.
To calculate the deviatoric part of the tensor $\DUinf \gdot\, \gSigma + \gSigma \gdot\, {\DUinf}^T$, we rewrite it as follows:

\begin{align}
	\DUinf \gdot\, \gSigma + \gSigma \gdot\, {\DUinf}^T &= \DUinf \gdot\, \gSigma' + \gSigma' \gdot\, {\DUinf}^T + \left(\tr\gSigma\right) \Einf \,. \nonumber\\
	\label{decomposition_duinfsigma}
\end{align}
The contribution proportional to $\Einf$ is traceless due to the incompressibility of the fluid. We calculate the deviatoric part of the remaining contributions:

\begin{align}
	&\left(\DUinf \gdot\, \gSigma' + \gSigma' \gdot\, {\DUinf}^T\right)' = \DUinf \gdot\, \gSigma' + \gSigma' \gdot\, {\DUinf}^T - \left(\Einf : \gSigma'\right)\identity \nonumber \\
	& \qquad \qquad \qquad \qquad = \Einf \gdot\, \gSigma' + \gSigma' \gdot\, \Einf + \Ominf \gdot\, \gSigma' - \gSigma' \gdot\, \Ominf  - \left(\Einf : \gSigma'\right)\identity
	\label{deviatoric_duinfsigmaprim}
\end{align}
The tensor $\gSigma'$ being symmetric, it can be diagonalized in an orthonormal basis $(\eone,\etwo)$. This tensor being traceless, it reads in this basis as

\begin{align}
\gSigma' = \mu \left(\eooeo-\etoet\right).
\end{align}
In the same basis, the tensor $\Einf$, which is traceless and symmetric, can be written as

\begin{align}
\Einf = \nu \left(\eooeo-\etoet\right)+\tilde{\nu} \left(\eooet+\etoeo\right).
\end{align}
The product of these two tensors thus reads:

\begin{align}
	\Einf\gdot\,\gSigma' &= \mu \nu \left(\eooeo+\etoet\right) + \mu \tilde{\nu} \left(-\eooet+\etoeo\right).
	\label{Einf_sigmaprim}
\end{align}
Moreover, we have in this notation $\Einf : \gSigma' = 2\mu \nu$, so that we get 

\begin{align}
\Einf \gdot\, \gSigma' + \gSigma' \gdot\, {\Einf}^T = \left(\Einf : \gSigma'\right)\identity.
\end{align}
Using all the above results in Eq.~(\ref{decomposition_duinfsigma}), we finally obtain Eq.~(\ref{deviatoric_duinfsigma}).

\section{\label{app:calc_integrals}Evaluation of the integrals involving \texorpdfstring{$\gthree$}{g3}}

In this Appendix, we evaluate the integrals involving $\gthree$ using the Kirkwood closure \eqref{kirkwood}.
We start by rewriting the expression (\ref{DL_g}) of $g(\R{})$ in a more convenient form for practical purposes.
The tensor $\gQ$ being symmetric and traceless, it can be diagonalized in an orthonormal basis $\left(\eone,\etwo\right)$ as
\begin{equation}
\gQ=\lambda\left(\eooeo-\etoet\right).
\end{equation}
Taking the direction $\eone$ as the origin for the polar angle $\theta$, $\er$ can be written as
\begin{equation}
\er=\cos\theta\,\eone+\sin\theta\,\etwo\,. 
\end{equation}
Eq.~(\ref{DL_g}) now reads:
\begin{equation}
	\label{expr_g_cos}
	g(\R{}) = \giso(r) + \lambda\alpha r \giso'(r)\cos2\theta.
\end{equation}
Using this expression we can rewrite the Kirkwood closure (\ref{kirkwood}) as a polynom of $\lambda$,

\begin{align}
	\label{dvpt_lambda_g3}
	\gthree(\R{},\rprim) = \gthree^{(0)}(\R{},\rprim) + \lambda \gthree^{(1)}(\R{},\rprim) + \lambda^2\gthree^{(2)}(\R{},\rprim) + \lambda^3\gthree^{(3)}(\R{},\rprim),
\end{align}
with the following functions,

\begin{align}
	\label{def_g30}
	& \gthree^{(0)}(\R{},\rprim) & = \giso(r)\giso(r')\giso(u)\,, \\
	\label{def_g31}
	&\gthree^{(1)}(\R{},\rprim)   &= \alpha r\,\giso'(r)\giso(r')\giso(u)\cos2\theta + \alpha r'\,\giso(r)\giso'(r')\giso(u)\cos2\theta' \nonumber\\
								  & &\quad + \alpha u\,\giso(r)\giso(r')\giso'(u)\cos2\phi\,,\\
	\label{def_g32}
	& \gthree^{(2)}(\R{},\rprim) &= \alpha^2 r r'\,\giso'(r)\giso'(r')\giso(u)\cos2\theta\cos2\theta' \nonumber\\
								   & &\quad + \alpha^2 r u\,\giso'(r)\giso(r')\giso'(u)\cos2\theta\cos2\phi \nonumber\\
								   & &\quad + \alpha^2 r' u\,\giso(r)\giso'(r')\giso'(u)\cos2\theta'\cos2\phi\,, \\
	\label{def_g33}
	& \gthree^{(3)}(\R{},\rprim) &= \alpha^3 r r' u\, \giso'(r) \giso'(r') \giso'(u) \cos2\theta\cos2\theta'\cos2\phi\,,
\end{align}
where $r=|\R{}|$, $r'=|\rprim|$, $u=|\R{}-\rprim|$, $\theta=\arg(\R{})$, $\theta'=\arg(\rprim)$ and $\phi=\arg(\R{}-\rprim)$. In addition, it is straightforward to show that

\begin{align}
	\label{cos2phi}
	u^2\cos2\phi=r^2\cos2\theta + r'^2\cos2\theta' - 2rr'\cos\left(\theta+\theta'\right),
\end{align}
an expression which will prove useful in what follows.

\subsection{Calculation of \texorpdfstring{$\gGamma$}{Gamma}}

We now evaluate the tensorial term $\gGamma$ defined in Eq.~\eqref{def_Gamma}.
We first note that its expression, defined by an integral over $(\R{},\rprim)$, can be simplified by considering symmetries of the integrand.
The tensorial part of the integrand reads
\begin{equation}
\gF(\R{})\otimes\gF(\rprim) = f(r)f(r')
\left(\begin{array}{ll}
\cos\theta\cos\theta' & \cos\theta\sin\theta' \\
\cos\theta'\sin\theta &\sin\theta\sin\theta'
\end{array}\right),
\end{equation}
so that diagonal terms are even and off-diagional terms are odd under the symmetry transformation
$(\theta,\theta')\rightarrow(-\theta,-\theta')$.
As $\gthree(\R{},\rprim)$ is invariant under this transformation, the off-diagonal terms of the integrand defining $\gGamma$ are odd under the transformation $(\theta,\theta')\rightarrow(-\theta,-\theta')$, so their integral is zero and $\gGamma$ is a diagonal tensor. Moreover, using the transformation $(\theta,\theta')\rightarrow(\frac{\pi}{2}-\theta,\frac{\pi}{2}-\theta')$ and the expression of the function $\gthree^{{(i)}}$, we can show that

\begin{align}
		\label{coscos_sinsin0}
		&\iint \cos\theta\cos\theta'\gthree^{(0)}(\R{},\rprim)\D\theta\D\theta' = \iint \sin\theta\sin\theta'\gthree^{(0)}(\R{},\rprim)\D\theta\D\theta', \\
		\label{coscos_sinsin1}
		&\iint \cos\theta\cos\theta'\gthree^{(1)}(\R{},\rprim)\D\theta\D\theta' = -\iint \sin\theta\sin\theta'\gthree^{(1)}(\R{},\rprim)\D\theta\D\theta', \\
		\label{coscos_sinsin2}
		&\iint \cos\theta\cos\theta'\gthree^{(2)}(\R{},\rprim)\D\theta\D\theta' = \iint \sin\theta\sin\theta'\gthree^{(2)}(\R{},\rprim)\D\theta\D\theta',\\
		\label{coscos_sinsin3}
		&\iint \cos\theta\cos\theta'\gthree^{(3)}(\R{},\rprim)\D\theta\D\theta' = -\iint \sin\theta\sin\theta'\gthree^{(3)}(\R{},\rprim)\D\theta\D\theta'.
\end{align} 
It follows that $\gGamma$ can be written as a polynomial in $\gSigma'$

\begin{equation}
	\label{expr_Gamma}
	\gGamma = \Gamma_0 \identity + \Gamma_1\gSigma' + \Gamma_2\gSigma'^2 + \Gamma_3\gSigma'^3,
\end{equation}
with $\gSigma'^2=\gSigma'\cdot\gSigma'$
and $\gSigma'^3=\gSigma'\cdot\gSigma'\cdot\gSigma'$,
and where the coefficients $\Gamma_i$ are given by

\begin{equation}
	\label{def_alphai}
	\Gamma_i = \frac{\rho^3}{2} \iint f(r)f(r') \cos\theta\cos\theta'\gthree^{(i)}(\R{},\rprim)\D\R{}\D\rprim.
\end{equation}
The deviatoric part of $\identity$ and $\gSigma'^{2}$ being null, while $\gSigma'$ and $\gSigma'^{3}$ are traceless, we can decompose $\gGamma$ as follows,

\begin{align}
	\label{deviatoric_Gamma}
	& \gGamma' = \Gamma_1 \gSigma' + \Gamma_3 \gSigma'^3, \\
	\label{tr_Gamma}
	& \tr\gGamma = 2 \Gamma_0 + \Gamma_2 \left(\gSigma':\gSigma'\right).
\end{align}
Replacing the functions $\gthree^{(i)}$ by their expressions, and using Eq.~(\ref{cos2phi}) to express $\cos2\phi$ as a function of $r$, $r'$, $\theta$ and $\theta'$, we can evaluate the coefficients $\Gamma_{i}$. Using trigonometric identities, it is possible to write all product of cosine fonctions as a sum of terms of form $\cos(n\theta+m\theta')$, with $n,m\in\mathbb{Z}$. In the case where $n+m\ne0$, $\cos(n\theta+m\theta')$ is odd under the $(\theta,\theta')\rightarrow(\theta+\pi/(n+m),\theta'+\pi/(n+m))$ transformation, whereas $u(\R{},\rprim)$ is conserved under this transformation. Thus the contribution of those terms to the integrals is zero. Only $\cos(n(\theta-\theta'))$ type terms have a non-zero contribution to the integrals.
Thus all integrands are only fonction of $\theta-\theta'$, so that we can reduce the angular variables number to one, performing $(\theta,\theta')\rightarrow(\theta-\theta',\theta')$ change of variables.
Noticing that $\gSigma'^{3}=\frac{1}{2}\left(\gSigma':\gSigma'\right)\gSigma'$, we obtain Eqs.~(\ref{expr_gammap}) and (\ref{expr_trgamma}), with the following expressions for the coefficients $\Gamma_{i}$,

\begin{align}
	\label{gamma0}
	& \Gamma_0 & = \pi\rho^3\int_0^2\int_0^2\int_{0}^{\pi} rf(r)\giso(r)r'f(r')\giso(r')\giso(u)\cos\theta\D r\D r'\D\theta, \\
	\label{gamma1}
	& \Gamma_1 & = \frac{\pi\alpha\rho^3}{2k} \left(2\int_0^2\int_0^2\int_{0}^{\pi} r^2f(r)\giso'(r)r'f(r')\giso(r')\giso(u)\cos\theta\D r\D r'\D\theta\right. \\ \nonumber
    									& & \left. + \int_0^2\int_0^2\int_{0}^{\pi} rf(r)\giso(r)r'f(r')\giso(r')\frac{\giso'(u)}{u}\left((r^2+r'^2)\cos\theta-2rr'\right)\D r\D r'\D\theta\right), \\
    \label{gamma2}
    & \Gamma_2 &=\frac{\pi\alpha^2\rho^3}{2k^2} \left(\int_0^2\int_0^2\int_{0}^{\pi} r^2f(r)\giso'(r)r'^2f(r')\giso'(r')\giso(u)\cos\theta\cos2\theta\D r\D r'\D\theta\right. \\ \nonumber
											& & \left. + 2\int_0^2\int_0^2\int_{0}^{\pi} rf(r)\giso(r)r'^2f(r')\giso'(r')\frac{\giso'(u)}{u}\left(r^2\cos2\theta+r'^2-2rr'\cos\theta\right)\D r\D r'\D\theta\right), \\
    \label{gamma3}
    & \Gamma_3 & = \frac{\pi\alpha^3\rho^3}{8k^3} \int_0^2\int_0^2\int_{0}^{\pi} r^2f(r)\giso'(r)r'^2f(r')\giso'(r') \\ \nonumber 
    & & \qquad\qquad\qquad\qquad\times\frac{\giso'(u)}{u}\left(1+2\cos2\theta\right)\left((r^2+r'^2)\cos\theta-2rr'\right)\D r\D r'\D\theta\,,
\end{align}
with $u=\sqrt{r^2+r'^2-2rr'\cos\theta}$

\subsection{Calculation of \texorpdfstring{$\gUpsilon$}{Upsilon}}

We now turn to the evaluation of the tensor $\gUpsilon$ defined by an integral in Eq.~\eqref{def_Upsilon}.
Using the expression (\ref{def_DF}) of $\nabla\gF$ in polar coordinates, we get

\begin{align}
	\label{rofDF}
	\left(\R{}\otimes\gF(\rprim)\right)\gdot \nabla\gF(\R{}) & = rf(r')f'(r)\cos(\theta'-\theta)\,\left(\eroer\right) \nonumber\\
    & \quad + f(r')f(r)\sin(\theta'-\theta)\,\left(\er\otimes\et\right).
\end{align}
Here again, the symmetries of the tensorial integrand simplify the evaluation of the integral.
The off-diagonals terms of the tensors $\sin(\theta'-\theta)\,\left(\er\otimes\et\right)\gthree(\R{},\rprim)$ and $\cos(\theta'-\theta)\,\left(\er\otimes\er\right)\gthree(\R{},\rprim)$ being odd under the transformation $(\theta,\theta')\rightarrow(-\theta,-\theta')$, $\gUpsilon$ is thus a diagonal tensor.
Moreover, using the transformation $(\theta,\theta')\rightarrow(\frac{\pi}{2}-\theta,\frac{\pi}{2}-\theta')$, we also obtain that

\begin{align}
	\label{sincossin0}
	&\iint \sin(\theta'-\theta)\cos\theta\sin\theta \,\gthree^{(0)}(\R{},\rprim) \D\theta\D\theta' = 0 \\
	\label{sincossin2}
	&\iint \sin(\theta'-\theta)\cos\theta\sin\theta \,\gthree^{(2)}(\R{},\rprim) \D\theta\D\theta' = 0.
\end{align}
Further using the transformation $(\theta,\theta')\rightarrow(\frac{\pi}{2}+\theta,\frac{\pi}{2}+\theta')$, we also show that
\begin{align}
	\label{sincossin1}
	&\iint \sin(\theta'-\theta)\cos\theta\sin\theta \,\gthree^{(1)}(\R{},\rprim) \D\theta\D\theta' = 0 \\
	\label{sincossin3}
	&\iint \sin(\theta'-\theta)\cos\theta\sin\theta \,\gthree^{(3)}(\R{},\rprim) \D\theta\D\theta' = 0,
\end{align}
so that the term $\sin(\theta'-\theta)\,\left(\er\otimes\et\right)$ does not contribute to $\gUpsilon$.

Then using the transformation $(\theta,\theta')\rightarrow(\frac{\pi}{2}-\theta,\frac{\pi}{2}-\theta')$ as well as the expressions of the functions $\gthree^{{(i)}}$, we finally obtain
\begin{align}
	\label{cos2_cos}
	&\iint \cos^2\theta\cos(\theta'-\theta)\gthree^{(0)}(\R{},\rprim)\D\theta\D\theta'  = \iint \sin^2\theta\cos(\theta'-\theta)\gthree^{(0)}(\R{},\rprim)\D\theta\D\theta', \nonumber\\ \\
	&\iint \cos^2\theta\cos(\theta'-\theta)\gthree^{(1)}(\R{},\rprim)\D\theta\D\theta' = -\iint \sin^2\theta\cos(\theta'-\theta)\gthree^{(1)}(\R{},\rprim)\D\theta\D\theta', \nonumber\\ \\
	&\iint \cos^2\theta\cos(\theta'-\theta)\gthree^{(2)}(\R{},\rprim)\D\theta\D\theta' = \iint \sin^2\theta\cos(\theta'-\theta)\gthree^{(2)}(\R{},\rprim)\D\theta\D\theta', \nonumber\\ \\
	&\iint \cos^2\theta\cos(\theta'-\theta)\gthree^{(3)}(\R{},\rprim)\D\theta\D\theta'  = -\iint \sin^2\theta\cos(\theta'-\theta)\gthree^{(3)}(\R{},\rprim)\D\theta\D\theta'. \nonumber\\ 
\end{align}

Performing the same reasoning as for $\gGamma$ from there, we get Eqs.~(\ref{expr_upsilonp}) and (\ref{expr_trupsilon}), with the following expressions for the coefficients $\Upsilon_{i}$,

\begin{align}
    \label{upsilon0}
    \Upsilon_0 & = \pi\rho^3\int_0^2\int_0^2\int_{0}^{\pi} r^2f(r)\giso(r)r'f'(r')\giso(r')\giso(u)\cos\theta\D r\D r'\D\theta, \\
    \label{upsilon1}
    \Upsilon_1 & =\frac{\pi\alpha\rho^3}{2k} \left(\int_0^2\int_0^2\int_{0}^{\pi} r^3f'(r)\giso'(r)r'f(r')\giso(r')\giso(u)\cos\theta\D r\D r'\D\theta\right. \\ \nonumber
    									& + \int_0^2\int_0^2\int_{0}^{\pi} r^2f'(r)\giso(r)r'^2f(r')\giso'(r') \giso(u)\cos\theta\cos2\theta\D r\D r'\D\theta \\ \nonumber
    									& + \int_0^2\int_0^2\int_{0}^{\pi} r^2f'(r)\giso(r)r'f(r')\giso(r')\frac{\giso'(u)}{u}\cos\theta \times \\ \nonumber
    								& \qquad \qquad \qquad \qquad \qquad \qquad \qquad \left.	\left(r^2+r'^2\cos2\theta-2rr'\cos\theta\right)\D r\D r'\D\theta\right), \\
    \label{upsilon2}
    \Upsilon_2 & =\frac{\pi\alpha^2\rho^3}{2k^2} \left(\int_0^2\int_0^2\int_{0}^{\pi} r^3f(r)\giso'(r)r'^2f'(r')\giso'(r')\giso(u)\cos\theta\cos2\theta\D r\D r'\D\theta\right. \\ \nonumber
											& + \int_0^2\int_0^2\int_{0}^{\pi} r^3f(r)\giso'(r)r'f'(r')\giso(r')\frac{\giso'(u)}{u}\left(r^2+r'^2\cos2\theta-2rr'\cos\theta\right)\D r\D r'\D\theta \\ \nonumber
											& \left. + \int_0^2\int_0^2\int_{0}^{\pi} r^2f(r)\giso(r)r'^2f'(r')\giso'(r')\frac{\giso'(u)}{u}\left(r^2\cos2\theta+r'^2-2rr'\cos\theta\right)\D r\D r'\D\theta\right), \\
    \label{upsilon3}
    \Upsilon_3 &=\frac{\pi\alpha^3\rho^3}{8k^3} \int_0^2\int_0^2\int_{0}^{\pi} r^3f'(r)\giso'(r)r'^2f(r')\giso'(r')\frac{\giso'(u)}{u}\cos\theta \\ \nonumber
        & \qquad \qquad \qquad \times \left(3r^2\cos2\theta+r'^2(2+\cos4\theta)-2rr'(2\cos\theta+\cos3\theta)\right)\D r\D r'\D\theta \,,
\end{align}
with $u=\sqrt{r^2+r'^2-2rr'\cos\theta}$.

\section{\label{app:exp_coef}Expression of the coefficients in terms of \texorpdfstring{$\rs$}{rs} and expansion in \texorpdfstring{$p$}{p}}

\subsection{Evaluation of the coefficients as a function of \texorpdfstring{$\rs$}{rs}}
\label{subsec:app:exp_coef:rs}

Using the parametrization (\ref{expr_giso}) of $\giso$, it is possible to evaluate all the coefficients $\Gamma_{i}$ and $\Upsilon_{i}$ defined in Eqs.~\eqref{expr_gammap}--\eqref{expr_trupsilon},
as well as the coefficients $\kappa$, $\beta$,\dots $\chi$ introduced in Eqs.~\eqref{def_A}--\eqref{def_F}. To make the resulting expressions tractable, we use one more approximation.
It turns out that the threefold product of functions $\giso$ and $\giso'$ in the integrals defining these coefficients make their exact calculation quite complicated. To make it simpler, we separate the angular part of the triple integrals from the ones on $r$ and $r'$ by making the approximation $r=r'=\rs$ in the angular part of the integrals. Doing so, we get for example $u\approx \sqrt{2}\rs\sqrt{1-\cos\theta}$. We obtain in this way the following expressions for the coefficients as a function of $\rs$,

\begin{align}
	\label{expr_A_rs}
	& \kappa = - \frac{\pi  \rho^{2}}{4} \rs \left(\rs^{3} - 2 \rs^{2} - 4 A \rs + 6 A\right), \\
	\label{expr_B_rs}
	& \beta = -\frac{2 \left(\rs^{4} - 3 \rs^{3} + \rs^{2} \left(2 - 4 A\right)+ 9 A \rs - 4 A\right)}{\rs \left( \rs^{3} - 2 \rs^{2} - 4 A \rs + 6 A\right)} - \Gamma_1 -\Upsilon_1, \\
	\label{expr_D_rs}
	& \zeta = -\frac{\rs^3-\rs^2-4A\rs+3A}{\rs^3-2\rs^2-4A\rs+6A}, \\
	\label{expr_E_rs}
	& \eta = - \frac{\pi \rho^{2}}{4}  \left(\rs - 2\right) \left(\rs^{3} - 2 \rs^{2}- 4 A \rs + 4 A\right) + \Gamma_0 + \Upsilon_0, \\
	\label{expr_Gamma0_rs}
	& \Gamma_0 = - \frac{\pi \rho^{3}}{18\sqrt{3} \rs^{2}} \left(\rs - 2\right)^{2} \left( 3 \rs^{2} - 2 A\right)  \left(\rs^{2} - \rs - 3 A - 2 \right)^{2}, \\
	\label{expr_Upsilon0_rs}
	& \Upsilon_0 = - \frac{\pi \rho^{3} }{48\sqrt{3} \rs^{2}} \left(\rs - 2\right) \left(3 \rs^{2} - 2A\right) \times \\ \nonumber
	& \qquad \qquad \left(\rs^{2} - 2A - 4\right) \left(3 \rs^{3} - 2 \rs^{2} - \rs \left(12 A + 4\right) - 8\right),	\\
	\label{expr_Gamma1_rs}
	& \Gamma_1 = \frac{2\rho\left(\rs-2\right)\left(\rs^2-\rs-3A-2\right)}{27\sqrt{3}\rs^3\left(\rs^3-2\rs^2-4A\rs+6A\right)} \times \\ \nonumber
& \qquad \qquad \left(21\rs^5-36\rs^4-73A\rs^3+(84A-24)\rs^2+30A^2\rs-24A^2+32A\right), \\
	\label{expr_Upsilon1_rs}
	&\Upsilon_1 = \frac{\rho}{54\sqrt{3}\rs^3\left(\rs^3-2\rs^2-4A\rs+6A\right)} \times \\ \nonumber
	& \qquad \qquad \Big[ 75 \rs^{8} - 198 \rs^{7} - 484 A \rs^{6} + \rs^{5} \left(1110 A + 24\right) + \rs^{4} \left(879 A^{2} + 288\right) \\ \nonumber
	& \qquad \qquad+ \rs^{3} \left(- 1788 A^{2} - 296 A\right) + \rs^{2} \left(- 306 A^{3} - 672 A + 192\right) \\ \nonumber
	& \qquad \qquad + \rs \left(720 A^{3} + 672 A^{2}\right) + 192 A^{2}  + 512 A\Big], \\
	\label{expr_Gamma2_rs}
	& \Gamma_2 = -\frac{2\left(3\rs^2 + 2A\right)\left(\rs+4\right)\left(\rs^2+2\rs-3A-8\right)\left(\rs^3-2\rs^2-3A\rs+4A\right)}{9\sqrt{3}\pi\rho\rs^4\left(\rs^3-2\rs^2-4A\rs+6A\right)^2}, \\
	\label{expr_Upsilon2_rs}
	& \Upsilon_2 = -\frac{4\left(3\rs^2 + 2A\right)}{9\sqrt{3}\pi\rho\rs^4\left(\rs^3-2\rs^2-4A\rs+6A\right)^2} \times \\ \nonumber
	& \qquad \qquad \left(\rs^5+12\rs^4-(5A+24)\rs^3-(48A+8)\rs^2+A(6A+72)\rs+16A\right), \\
	\label{expr_Gamma3_rs}
	& \Gamma_3 = -\frac{64A\left(\rs^3-2\rs^2-3A\rs+4A\right)^2}{\sqrt{3}\pi^2\rho^3\rs^5\left(\rs^3-2\rs^2-4A\rs+6A\right)^3}, \\
	\label{expr_Upsilon3_rs}
	& \Upsilon_3 = -\frac{16A\left(\rs^2-3A\right)\left(\rs^3-2\rs^2-3A\rs+4A\right)}{\sqrt{3}\pi^2\rho^3\rs^4\left(\rs^3-2\rs^2-4A\rs+6A\right)^3}.
\end{align}
Note that the coefficients $\xi$ and $\chi$ are expressed as simple combinations of $\Gamma_i$ and $\Upsilon_i$
[see Eqs.~\eqref{def_C} and \eqref{def_F}] so that they are not displayed explicitly here.
For completeness, we also give the explicit expression of the proportionality coefficient $k$ linking the deviatoric stress tensor $\gSigma'$ to the structure tensor $\gQ$,
\begin{equation}
    k = \frac{\rs^3 - 2\rs^2-4A\rs+6A}{\rs (\rs^2-4A)}\,.
\end{equation}

\subsection{Expansion in powers of \texorpdfstring{$p$}{p}}

The pressure being a macroscopic observable, it may be desirable to keep the variable $p$ in the description instead of $\rs$.
However, the relation $p(\rs)$ given in Eq.~(\ref{expr_p_rs}) cannot be easily inverted analytically, and one would need to resort to numerical methods. Yet, it can be inverted in a perturbative way for small $p$, corresponding to $\rs \apprle 2$. Expanding Eq.~(\ref{expr_p_rs}) to second order in $\varepsilon=2-\rs$, we obtain

\begin{equation}
	\label{DL_p_epsilon}
	p \approx 3\rho\varepsilon + \frac{\pi\rho^2(2-A)}{2}\varepsilon^2 + o(\varepsilon^2)\,.
\end{equation}
Perturbatively evaluating the inverse relation $\rs(p)$, we get to order $p^2$

\begin{equation}
	\label{DL_epsilon_p}
	\varepsilon \approx \frac{p}{3\rho} + \frac{2-A}{2\pi^2\rho^4A^3}p^2 + o(p^2)\,.
\end{equation}
Using the results of \ref{subsec:app:exp_coef:rs},
we can expand to second order in $\varepsilon$ the expressions of the coefficients $\kappa$, $\beta$, \dots $\chi$, and use Eq.~(\ref{DL_epsilon_p}) to obtain the expansion in $p$ of these coefficients up to order $p^{2}$.
One then finds

\begin{align}
	\label{expr_A_p}
	& \kappa = \pi A\rho^2 - \frac{5 A - 4}{2A} \, p - \frac{4 \pi^{2} A^{4} \rho^{2} - 12 \pi^{2} A^{3} \rho^{2} - 45 A^{2} + 126 A - 72}{36 \pi A^{3} \rho^{2}} p^2 +o(p^2)\,, \\
	\label{expr_B_p}
	&\beta = \frac{\rho A(A+6)}{4\sqrt{3}} - 1 
		  + \left(\frac{47 \sqrt{3} A}{72} - \frac{9 \sqrt{3}}{4} + \frac{3\sqrt{3}}{A} + \frac{1}{A \rho}\right) \frac{p}{\pi\rho} \nonumber\\
		  &\quad + \left(\frac{A \left(247 \sqrt{3} A^{2} \rho - 572 \sqrt{3} A \rho + 432 \sqrt{3} \rho + 216\right) - 288 \sqrt{3} \rho - 144}{1296 A \rho^2} \right. \nonumber\\ 
		  &\qquad\qquad\qquad+ \left. \frac{9 \left(A - 2\right) \left(47 \sqrt{3} A^{2} \rho - 162 \sqrt{3} A \rho + 216 \sqrt{3} \rho + 72\right)}{1296 \pi^{2} A^{3} \rho^{4}}\right) p^2 +o(p^2)\,, \\
	\label{expr_C_p}
	&\xi  = \frac{7A-4}{8\sqrt{3} \pi^2A\rho^3 } + \frac{83A^2-84A+32}{16\sqrt{3}\pi^3A^3\rho^5 } p \nonumber\\
		  &\quad+ \left(\frac{630 A^{3} - 936 A^{2} + 640 A - 192}{288\sqrt{3}\pi^2 A^3 \rho^5} + \frac{\left(A - 2\right) \left(83 A^{2} - 84 A + 32\right)}{32\sqrt{3}\pi^{4} A^{5} \rho^{7}}\right) p^2 +o(p^2)\,, \\
	\label{expr_D_p}
	&\zeta  = \frac{1}{2 A} \left(4-5A\right) - \frac{3A^2-5A+4}{\pi A^3\rho^2} p \nonumber\\
		  & \quad - \left(\frac{ 12 A^{3} - 27 A^{2} + 28 A - 16}{18 A^3 \rho^2} + \frac{ \left(A - 2\right) \left(3 A^{2} - 5 A + 4\right)}{2 \pi^{2} A^{5} \rho^{4}}\right) p^2 +o(p^2)\,, \\ \nonumber\\ \nonumber\\ \nonumber\\
	\label{expr_E_p}
	&\eta = - \left(\frac{\rho A(A-6)}{2 \sqrt{3}} + 1 \right) p \nonumber\\
		  & \quad - \left(\frac{\pi\left( 6 \sqrt{3} A^{2} \rho - 12 \sqrt{3} A \rho - 4 A + 4\right)}{36} + \frac{\left(A - 2\right) \left(\sqrt{3} A \rho \left(A - 6\right) + 6\right)}{12\pi A^{2} \rho^{2}}\right) p^2 +o(p^2)\,, \\
	\label{expr_F_p}
	&\chi = \frac{(A+6)(12-5A)}{24\sqrt{3}\pi A\rho} - \frac{6A^3+16A^2-51A+36}{6\sqrt{3}\pi^2A^3\rho^3} p \nonumber\\
		  &\quad -\left(\frac{103 A^{4} + 216 A^{3} - 768 A^{2} + 800 A - 384}{288\sqrt{3} \pi A^3 \rho^3} + \frac{\left(A - 2\right) \left(6 A^{3} + 16 A^{2} - 51 A + 36\right)}{12\sqrt{3} \pi^{3} A^{5} \rho^{5}}\right) p^2 +o(p^2)\,. \nonumber\\
\end{align}

Using the first-order expansion in $p$ of these coefficients, we can solve the equation (\ref{stationnary_p}) and get the following approximate expression for the stationary value of $p$ close to jamming (where we replaced $A$ by $3/\phi$ and $\rho$ by $\phi/\pi$) at first-order in $\phi-\phi_{\rm J}$,

\begin{align}
    \label{approximated_p_phi}
    p\approx\frac{315}{2\pi}\times\frac{21\sqrt{3}-10\pi}{753\sqrt{3}-290\pi}(\phi-\phi_{\rm J}).
\end{align}

Replacing $p$ in the pressure expansion of coefficients $\kappa$, $\beta$ and $\xi$ we get at first-order in $\phi-\phi_{\rm J}$

\begin{align}
    \label{approximated_bar_kappa_phi}
    \bar{\kappa} & \approx \frac{15}{4\pi}+\frac{27(130\pi-241\sqrt{3})}{2\pi(753\sqrt{3}-290\pi)} (\phi-\phi_{\rm J}) \\
    \label{approximated_bar_beta_phi}
    \bar{\beta} & \approx\frac{21\sqrt{3}}{10\pi}-1+\frac{3\left(-28449 - 2560 \sqrt{3} \pi + 3500 \pi^{2}\right)}{25\pi \left(- 753 \sqrt{3} + 290 \pi\right)}\left(\phi - \phi_{\rm J}\right) \\
    \label{approximated_bar_xi_phi}
    \bar{\xi} & \approx\frac{128 \sqrt{3} \pi}{1125} + \frac{16 \pi \left(-97317 + 17870 \sqrt{3} \pi\right)}{1875 \left(- 753 \sqrt{3} + 290 \pi\right)}\left(\phi - \phi_{\rm J}\right)
\end{align}

\section{\label{app:temp}Inclusion of a small thermal noise}

\subsection{Derivation of the equation on \texorpdfstring{$\gtwo$}{gtwo}}

In this Appendix, we provide a detailed evaluation of the additional term $\gPsi$ arising from the inclusion of a very small temperature in the dynamics. The dynamics remains essentially athermal and dominated by elastic forces.
The dimensionless equation for the particle dynamics now takes the form
\begin{equation}
	\label{dynamics:app:temp}
	-2 (\dot{\bm{r}}_\mu - \bm{u}^{\infty}_{\mu})+\bm{f}_{\mu} + \bm{\eta}_{\mu}(t)= \bm{0}\,,
\end{equation}
where $\bm{\eta}_{\mu}(t)$ is a Gaussian white noise with correlation
\begin{equation}
    \langle \eta_{\mu,i}(t)\eta_{\nu,j}(t')\rangle = 4 T \delta_{ij} \, \delta_{\mu\nu} \, \delta(t-t')\,.
\end{equation}
Note that the dimensionless temperature is expressed in units of $T_0=\frac{a f_0}{k_B}$.
The density current $\op{j}{}{\mu}$ associated with particle $\mu$ reads
\begin{align}
	\label{j_alpha:app:temp}
	\op{j}{}{\mu}=\left(\op{u}{\infty}{\mu} + \frac{1}{2}\op{f}{}{\mu}\right)P_N - \frac{T}{2} \nabla_\mu P_N\,.
\end{align}
Following similar steps as in Sec.~\ref{sec:level1}, one finds for the evolution equation for $\gtwo$,
\begin{equation}
	\label{eq_g2:app:temp}
	\partial_t \gtwo(\R{}) + \mathbf{\nabla} \gdot\left(\DUinf \gdot \R{}\,\gtwo(\R{}) -\gF(\R{})\gtwo(\R{})-\rho \int \gF(\rprim) \gthree(\R{},\rprim) \D \rprim\right) - T \nabla^2 \gtwo(\R{}) = 0,
\end{equation}
Multiplying the latter equation by $\rho^2\left(\rof\right)/2$ and integrating over $\bm{r}$, one finds the following tensorial equation
\begin{equation}
	\label{eq_sigma:app:temp}
	\dot{\gSigma} = \DUinf \gdot\, \gSigma + \gSigma \gdot\, {\DUinf}^T + \mathbf{\Theta} - \mathbf{\Phi} - \mathbf{\Xi} - \mathbf{\Pi} - \mathbf{\Gamma} - \mathbf{\Upsilon} + \mathbf{\Psi},
\end{equation}
with $\gPsi=\frac{\rho^2 T}{2}\int\left(\rof\right)\nabla^2 g(\R{})\D\R{}$.

\subsection{Expression of \texorpdfstring{$\mathbf{\Psi}$}{Psi}}

Using the Green-Ostrogradski formula, one finds (again for a radial force $\gF$)
\begin{align}
	\label{step1:app:temp}
	\gPsi=-\frac{\rho^2 T}{2}\int\nablar g\otimes\gF\D\R{} - \frac{\rho^2 T}{2}\int \left(\R{}\otimes\nablar g\right)\gdot\nablar\gF\, \D\R{}
\end{align}
Applying once more the same formula, one obtains
\begin{align}
	\label{step2:app:temp}
	\int\nablar g\otimes\gF\D\R{}= - \int \nablar\gF \, \gtwo(\R{})\D\R{},
\end{align}
as well as
\begin{align}
	\label{step3:app:temp}
	\int \! \left(\R{}\otimes\nablar\gF \right)\!\gdot\!\nablar g\, \D\R{} = \! \int_S \gtwo(\R{})\left(\R{}\otimes\!\nablar\gF\gdot\D\mathbf{S}\right) - \! \int \nablar\gF \gtwo(\R{})\D\R{} - \!\int \left(\R{}\otimes\nabla^2 \gF\right)\gtwo(\R{})\D\R{},
\end{align}
We thus have
\begin{align}
	\label{exp_Psi:app:temp}
	\gPsi = \frac{\rho^2T}{2}\left(-\int_S \gtwo(\R{})\left(\R{}\otimes\nabla\gF\gdot\D\mathbf{S}\right) + 2 \int \nabla\gF \,\gtwo(\R{})\D\R{} +\int \left(\R{}\otimes\nabla^2 \gF\right)\gtwo(\R{})\D\R{}\right)
\end{align}
with
\begin{align}
	\label{def_DF:app:temp}
		\nabla\gF = f'(r) (\er\otimes\er) + \frac{f(r)}{r} (\et\otimes\et) 
\end{align}
and
\begin{align}
	\label{def_laplacienF:app:temp}
		\nabla^2 \gF = \left(f''(r)+\frac{f'(r)}{r}-\frac{f(r)}{r^2}\right)\er
\end{align}

\subsection{Expression of \texorpdfstring{$\mathbf{\Psi}$}{Psi} as a function of \texorpdfstring{$\mathbf{\Sigma}$}{Sigma}}

We start by evaluating $\int_S \gtwo(\R{})\left(\R{}\otimes\nablar\gF\gdot\D\mathbf{S}\right)$.
Replacing $\nablar\gF$ by its expression \eqref{def_DF:app:temp}, one has:
\begin{align}
	\label{I1_step1:app:temp}
	\int_S \gtwo(\R{})\left(\R{}\otimes\nablar\gF\gdot\D\mathbf{S}\right)=4f'(2)\int_{-\pi}^{\pi} \gtwo(2,\theta)\left(\eroer\right)\D\theta
\end{align}
Using the parametrization \eqref{DL_g} of $\gtwo$ as well as the integral properties given in Eqs.~(\ref{integralg_prop1}--\ref{integralg_prop5}), we then obtain
\begin{align}
	\label{I1_final:app:temp}
	\int_S \gtwo(\R{})\left(\R{}\otimes\nablar\gF\gdot\D\mathbf{S}\right)=4\pi f'(2)\giso(2)\identity +\frac{4\alpha\pi f'(2)\giso'(2)}{k}\gSigma'
\end{align}
We then turn to the evaluation of $\int \nablar\gF \gtwo(\R{})\D\R{}$.
Replacing $\nablar\gF$ by its expression \eqref{def_DF:app:temp}, we get
\begin{align}
	\label{I2_step1:app:temp}
	\int \nablar\gF \,\gtwo(\R{})\D\R{} = \int f'(r) (\er\otimes\er)\,\gtwo(\R{})\D\R{} + \int\frac{f(r)}{r} (\et\otimes\et)\,\gtwo(\R{})\D\R{}
\end{align}
Using again the parametrization \eqref{DL_g} of $\gtwo$ together with Eqs.~(\ref{integralg_prop1}--\ref{integralg_prop5}), one obtains
\begin{align}
	\label{I2_step2:app:temp}
	\int f'(r) (\er\otimes\er)\,\gtwo(\R{})\D\R{} = \pi\left(\int_0^2 rf'(r)\giso(r)\D r\right)\identity + \frac{\pi\alpha}{2k}\left(\int_0^2r^2f'(r)\giso'(r)\D r\right)\gSigma'
\end{align}
and
\begin{align}
	\label{I2_step3:app:temp}
	\int\frac{f(r)}{r} (\et\otimes\et)\,\gtwo(\R{})\D\R{} = \pi\left(\int_0^2 f(r)\giso(r)\D r\right)\identity - \frac{\pi\alpha}{2k}\left(\int_0^2rf(r)\giso'(r)\D r\right)\gSigma'
\end{align}
We finally determine the integral $\int \left(\R{}\otimes\nabla^2 \gF\right)\gtwo(\R{})\D\R{}$.
We replace $\nabla^2 \gF$ by its expression \eqref{def_laplacienF:app:temp}, and $\gtwo$ by its parametrization \eqref{DL_g}. Using again Eqs.~(\ref{integralg_prop1}--\ref{integralg_prop5}), we obtain
\begin{align}
	\label{I3_step1:app:temp}
	\int \left(\R{}\otimes\nabla^2 \gF\right)\gtwo(\R{})\D\R{} & = \pi\left(\int_0^2\left(r^2f''(r)+rf'(r)-f(r)\right)\giso(r)\D r\right)\identity \\
								& + \frac{\alpha\pi}{2k}\left(\int_0^2r\left(r^2f''(r)+rf'(r)-f(r)\right)\giso'(r)\D r\right)\gSigma'
\end{align}
Altogether, we finally get
\begin{align}
	\label{exp_Psi_Sigma:app:temp}
	\gPsi = &\frac{\pi\rho^2T}{2}\left(\int_0^2 \left(r^2f''(r)+3rf'(r)+f(r)\right)\giso(r)\D r-4 f'(2)\giso(2)\right)\identity \nonumber \\
	 	  + &\frac{\alpha\pi\rho^2T}{4k}\left(\int_0^2r\left(r^2f''(r)+3rf'(r)-3f(r)\right)\giso'(r)\D r-8f'(2)\giso'(2)\right)\gSigma'
\end{align}
Choosing $f(r)=r-2$, $\giso(2)=1$ and $\giso'(2)=0$, this expression becomes
\begin{align}
	\label{exp_Psi_Sigma_simplif:app:temp}
	\gPsi = \pi\rho^2T\left(\int_0^2 (2r-1)\giso(r)\D r-2 \right)\identity + \frac{3\alpha\pi\rho^2T}{2k}\left(\int_0^2r\giso'(r)\D r\right)\gSigma'
\end{align}
Replacing $\giso$ by its parametrization $\giso(r)=\frac{A}{\rs}\delta(r-\rs)+\Theta(r-\rs)$, we obtain the following expression in terms of $\rs$,
\begin{align}
	\label{exp_Psi_Sigma_rstar:app:temp}
	\gPsi = -\pi\rho^2T \, \frac{\rs^3-\rs^2-2A\rs+A}{\rs} \, \identity + 6T\frac{\rs^2-A}{\rs^2\left(\rs^3-2\rs^2-4A\rs+6A\right)}\gSigma'
\end{align}
Expanding this expression to first order in the pressure $p$ eventually leads to
\begin{align}
	\label{exp_Psi_p:app:temp}
	\gPsi = \pi\rho^2T\left(\frac{3A-4}{2} + \frac{12-A}{4\pi A\rho^2}p \right)\identity + 3T\left(\frac{A-4}{4A}+\frac{(A-2)(3A-4)}{4\pi A^3\rho^2}p\right)\gSigma'
\end{align}

\section{\label{app:evol_p}Behaviour of the model with dynamics of the pressure}

\begin{figure}[b]
	\begin{center}
		\includegraphics[width=0.32\linewidth]{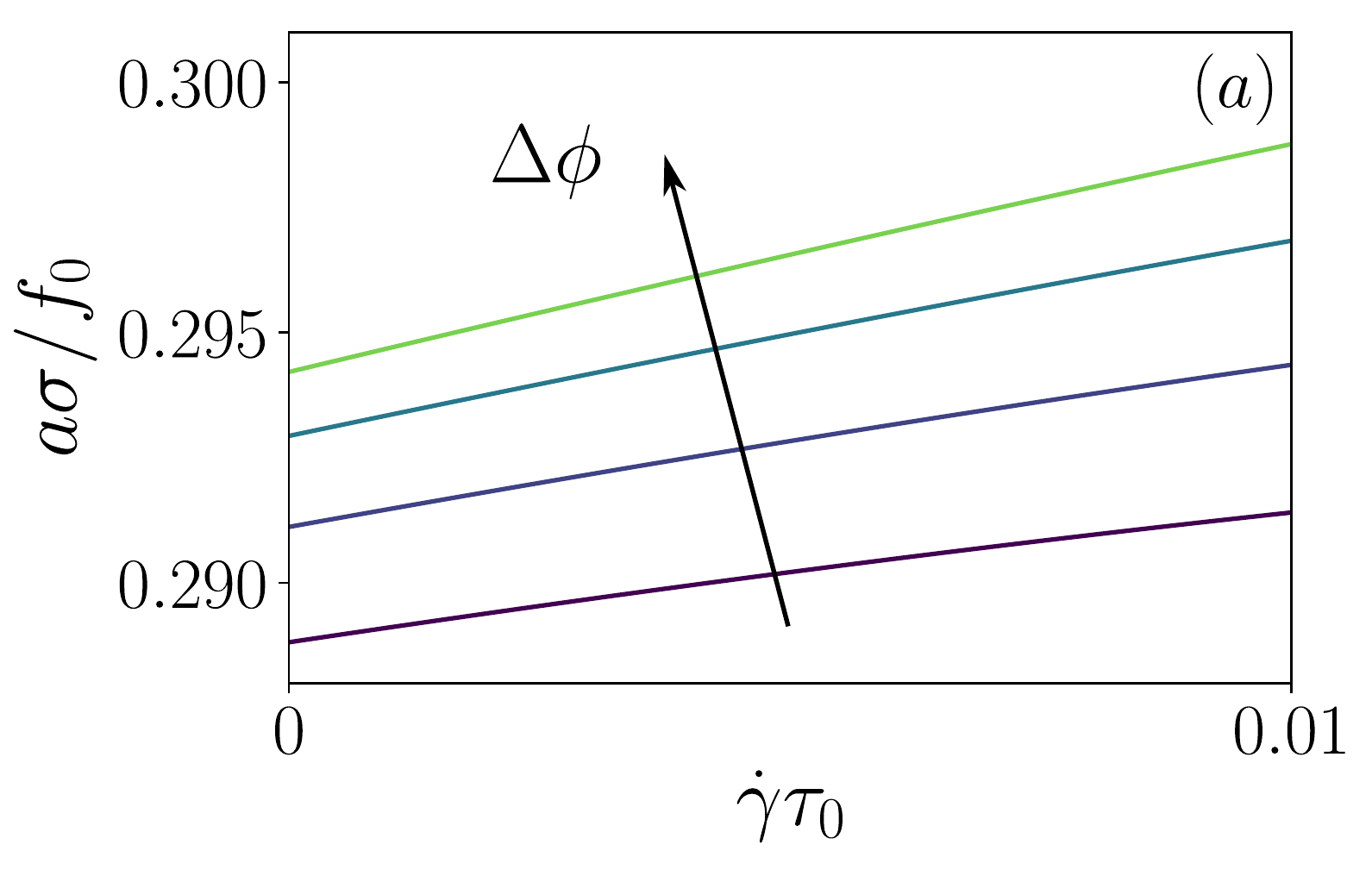}
		\includegraphics[width=0.32\linewidth]{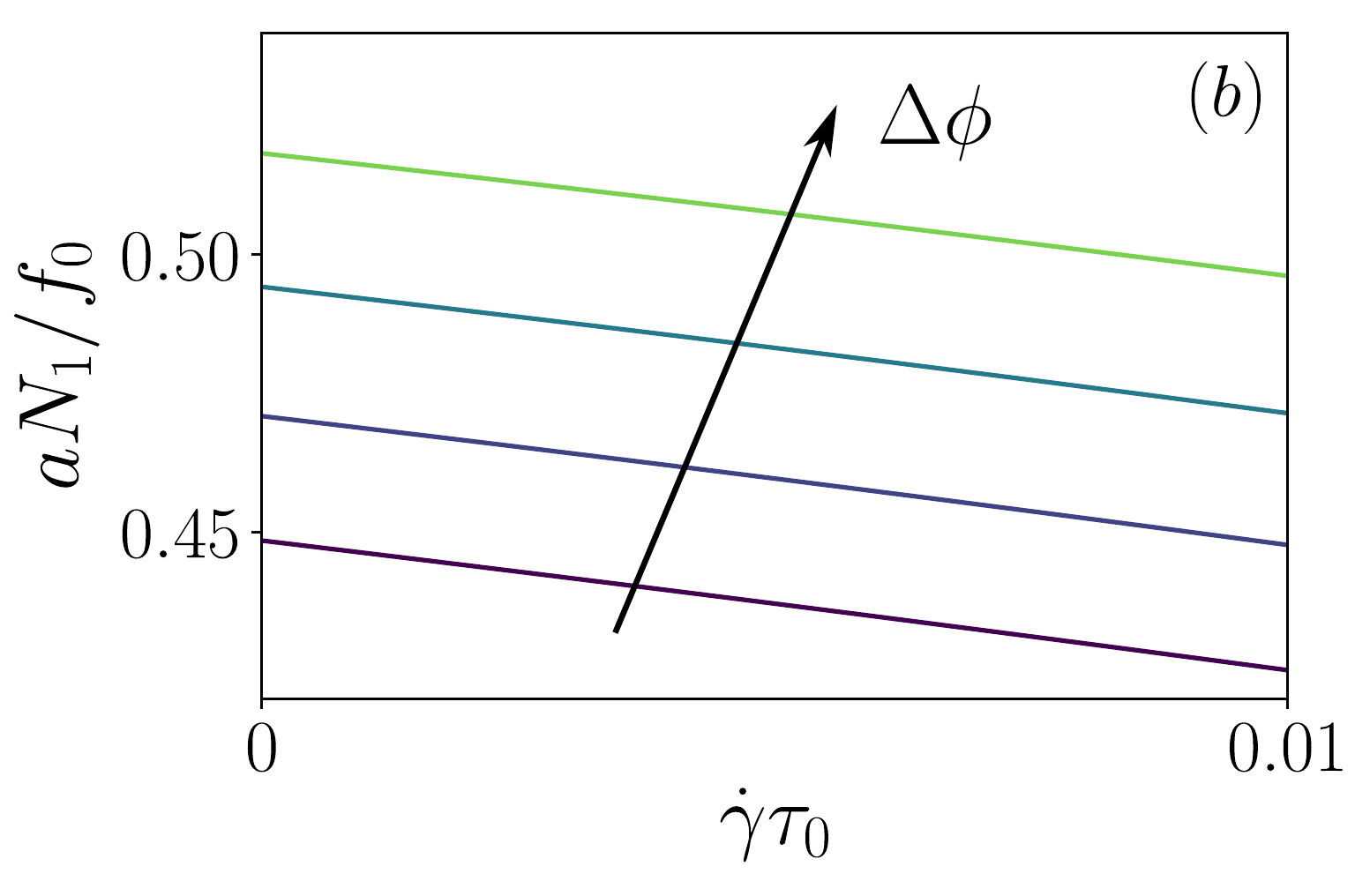}
		\includegraphics[width=0.32\linewidth]{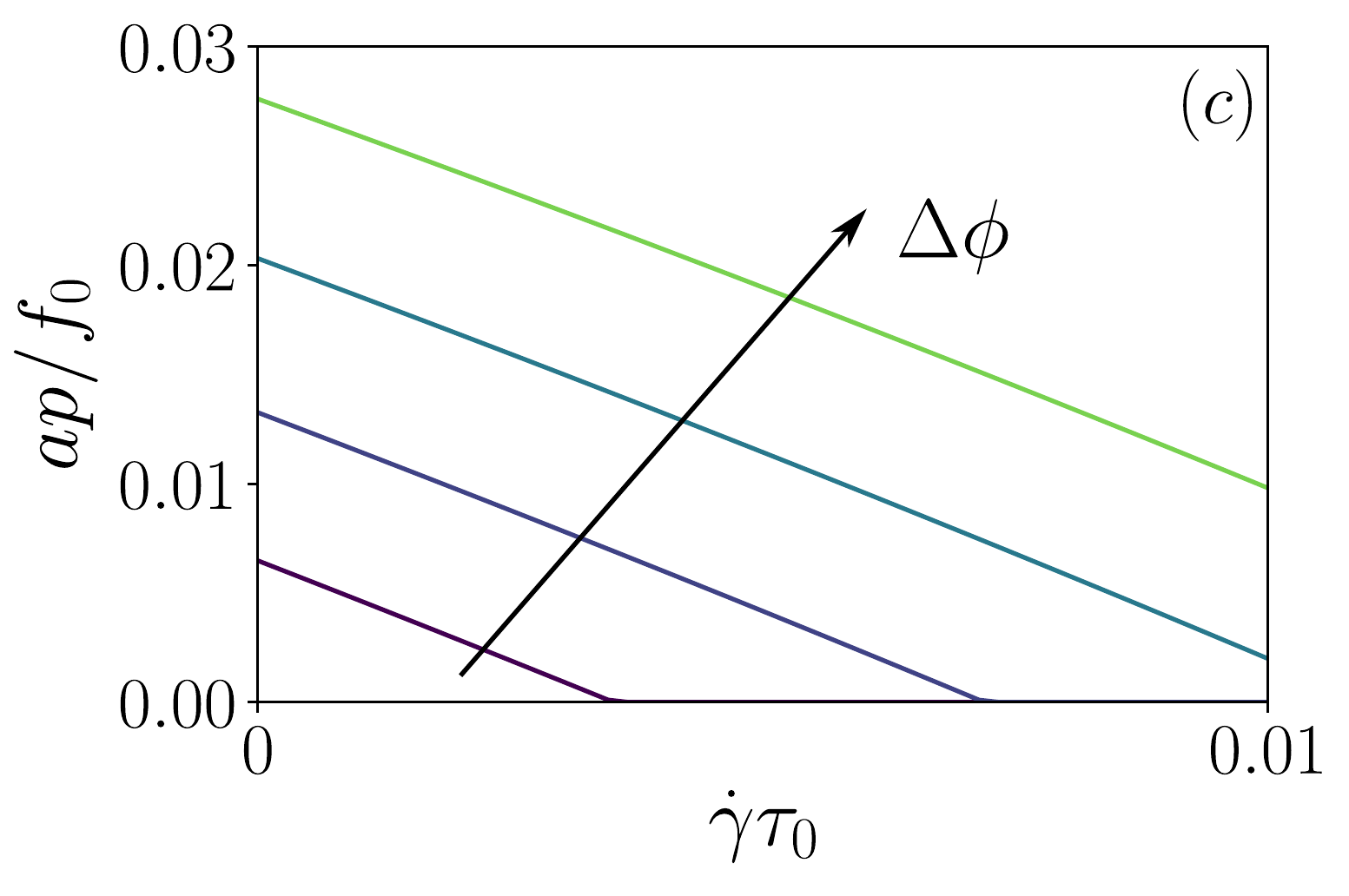}
		\caption{\comnc{Evolution of the shear stress (a), the normal stress difference (b) and the pressure (c) for different values of surface fraction ($\Delta\phi=1\times10^{-2},\,2\times10^{-2},\,3\times10^{-2},\,4\times10^{-2}$) under the coupled equations (\ref{eq_sigmaprim_closed_p}) and (\ref{eq_p_closed_p}).
		}}
		\label{fig:evol_p}
	\end{center}
\end{figure}

\comnc{In this Appendix we briefly present the behaviour of our model for a simple shear deformation in case we solve the full system of Eqs.~\eqref{eq_sigmaprim_closed_p} and \eqref{eq_p_closed_p} instead of assuming $p$ to be constant by using a stationary value of the pressure depending of the surface fraction only. We show in Fig.~\ref{fig:evol_p} the flow curves on $\sigma$, $N_1$ and $p$ obtained for different values of $\phi$. If $\sigma$ still exhibits a yield-stress and grows with the shear rate, we observe that $N_1$ and $p$ decrease with the shear rate, which is an unexpected behaviour. The pressure can even become zero at high shear rate for system close enough to jamming. We explain this unexpected behaviour of the pressure by the fact that the parametrization of the isotropic part of the pair correlation function we choose (Eq.~\eqref{expr_giso}) is not able to conserve the correct dependence of the pressure on the anisotropy of the microstructure. Indeed with this parametrization the first-neighbour shell is a circle of radius $r^*$. In this framework the pressure is insensitive to the anisotropy of the microstructure [measured by the norm of the structure tensor, $|\bm{Q}|=\sqrt{(\bm{Q}:\bm{Q})/2}$] and vary only with $r^*$. However, it has been observed in some numerical studies of sheared dense suspensions that the growth of anisotropy of the system during the shear is accompanied by an accumulation of contacts along the compressional axis and a depletion of contacts along the elongational axis, resulting in a pressure growth
\cite{mohan_microscopic_2013,mohan_build-up_2014}. 
As we do not have the correct qualitative behaviour for the pressure with respect to the anisotropy we assume that the equation on the pressure we obtain is not able to give the right qualitative dependence of the pressure with the shear rate.}

\vspace*{5mm}
% \nocite{*}

\bibliographystyle{plain_url.bst}
\bibliography{biblio}% Produces the bibliography via BibTeX.

\end{document}